\documentclass{astlb}

\usepackage{apjfonts}
\usepackage{amsmath}
\usepackage{epsfig,graphics}
\usepackage{rotating} 
\usepackage{epsfig}

\usepackage{graphicx,natbib,multirow,topcapt}
\usepackage{savefnmark}
\usepackage[perpage,symbol*]{footmisc}
\bibpunct{(}{)}{;}{a}{}{,}  
\newcommand {\be}{\begin{equation}}
\newcommand {\ee}{\end{equation}}  

\def\smfigure#1#2#3{
  \centering	
  \begin{minipage}{1.\columnwidth}
    \begin{minipage}{0.049\columnwidth}
      \rotatebox{90}{\small\phantom{0000}#3}
    \end{minipage}
    \begin{minipage}{1.\columnwidth}   
     
      \centerline{\small #2}
      \includegraphics[bb=40 208 556 678,width=0.9\columnwidth]{#1}
     
    \end{minipage}

    \vskip 30pt
    ~
  \end{minipage}
}

\begin{document}
\journalinfo{2012}{38}{7}{443}[467] 

\title{X-ray line formation in the spectrum of SS 433 }
\author{\bf %\hspace{-1.3cm} \copyright 
I.I. Khabibullin \email{khabibullin@iki.rssi.ru}\address{1},
S.Yu. Sazonov
\address{1}
\addresstext{1}{Space Research Institute, Russian Academy of Sciences, Profsoyuznaya Str. 84/32, Moscow, 117997 Russia}
}
\shortauthor{Khabibullin \& Sazonov}
\shorttitle{X-ray line formation }
\submitted{Dec. 2, 2011}
%\onecolumn
\begin{abstract}
 The mechanisms for the formation of X-ray lines in the spectrum of SS 433 are investigated by taking into account the radiative transfer inside the jets. The results of Monte Carlo numerical simulations are presented. The effect of a decrease in line intensity due to scattering inside the jet turns out to be pronounced, but it does not exceed 60 \% in magnitude on the entire grid of parameters. The line broadening due to scattering, nutational motion, and the contribution of satellites can lead to overestimates of the
jet opening angle $\Theta  $ from the line widths in Chandra X-ray observations. The fine structure of the lines turns out to be very sensitive to the scattering effects. This makes its investigation by planned X-ray observatories equipped with high-resolution spectrometers (primarily Astro-H) a powerful tool for diagnosing the parameters of the jets in SS 433.

% 
%\medskip
%PACS: 
%95.55.Aq, 95.55.Cs, 95.75.De, 95.75.Wx, 95.85.Kr, 97.30.Sw, 97.60.Lf, 97.80.Jp
\end{abstract}

\section{INTRODUCTION}

\ \ \ \ The Galactic source SS 433 is currently the only
close X-ray binary system observed in a permanently
supercritical regime of accretion. This manifests itself
most clearly in the presence of a pair of oppositely
directed relativistic jets oriented perpendicular to the
plane of a supercritical accretion disk (for a review, see
\cite{fabrika04}). \
 
	X-ray observations of the system show that apart
from the continuum component attributable mainly
to bremsstrahlung from the jets, its spectrum exhibits
numerous lines of highly ionized atoms of heavy
elements. By measuring their characteristics, one
can get an idea of the physical conditions in the
regions closest to the compact object in which the
jets are formed, accelerated, and collimated 	\citep{kotani96}. \

Substantial progress in understanding these fundamental
mechanisms has been achieved in the
last ten years through observations by the Chandra \citep{marshetal02,nametal03,lopezetal06}, XMM-Newton \citep{brinketal05},
INTEGRAL\citep{cher05}, RXTE
\citep{kateetal06}, and Suzaku \citep{suzaku10}
X-ray observatories in the standard
and hard spectral ranges. However, the uncertainty in
physical parameters of the jets (density, size, and
opening angle) remains significant (for more details,
see Sections 4.3 and 6). With high-spectral resolution
instruments (primarily Chandra HETGS),
it has been possible to measure the line widths and
intensities and to estimate the opening angle, density,
and temperature of the hottest parts of the jets from
them (\cite{marshetal02},\cite{nametal03}, \cite{lopezetal06}). 
However, these required correcting
the previously adopted standard jet model \citep{kotani96}. 
For example, the emission measure of
the hottest parts of the jet estimated from the line
intensity in the high-energy part of the spectrum
turned out to be approximately half the value expected
from the standard model. In addition, it was
required to assume an overabundance (about 30\%
relative to the solar value) of iron, sulfur, silicon,
magnesium, and neon \citep{marshetal02} for the
observed continuum to agree with that predicted
from the line intensities. The EPIC/XMM-Newton
spectra point to a more than eightfold excess of
radiation near the $ K_\alpha$ triplet of helium-like nickel
(\cite{brinketal05},\cite{medved10}).
The physical interpretation of such discrepancies
was often reduced basically either to the uniqueness
of the SS 433 phenomenon or to the inclusion of
additional emitting or absorbing components of the
system while restricting oneself to using a very simple
scenario for the formation of X-ray lines. \

  In this paper, we study in detail the influence of
scattering effects inside the jets on the spectrum of
emergent radiation in lines of the standard X-ray band
for the first time. The necessity of such a study follows
from simple estimates of the jet transverse optical
depth for Thomson scattering ($ \sim 0.1 $) and, what is especially
important, for resonant scattering on electron transitions 
in highly ionized atoms of heavy elements
($ \sim 100 $ for the permitted $ K_{\alpha} $  transitions in helium-like
ions). As a consequence, the effect of an increase in
the mean free path of a photon before its escape from
the jet due to multiple resonant scatterings becomes
important. This, in turn, leads to an increase in
the probability of being scattered by a ``hot'' electron
and leaving the line while reducing its intensity and
forming broad wings.

Using the Monte Carlo technique for radiative
transfer problems developed by \cite{pozetal83}
allowed the formation of X-ray lines
in the spectrum of the jets to be simulated most
accurately by taking into account the multiplicity of
the most important of them and the possible deviations
from the coronal approximation. As a result,
we made predictions regarding the observational line
characteristics, bearing in mind the high resolution of
planned spectroscopic experiments (primarily Astro-H; 
see, e.g., \cite{takahashi10}.

%%%%%%%%%%%%%%%%%%%%%%%%%%%%%%%%%%%%%%%%%%%%%%
 \section{THE JET MODEL}
%%%%%%%%%%%%%%%%%%%%%%%%%%%%%%%%%%%%%%%%%%%%%%
\ \ \ \  The standard multitemperature jet model for
SS 433 \citep{kotani96, marshetal02}
served as a starting point for our study. In this model,
the X-ray jet is treated as a highly collimated (an opening
angle $ \Theta\sim 1^{\circ} $ ) ballistic plasma flow directed
away from the compact object perpendicular to the
plane of a supercritical accretion disk. The velocity
component parallel to the jet axis is constant at each
point and equal to 0.26c. The matter is distributed
uniformly within each layer perpendicular to the jet
axis. Under the assumption of an axisymmetric
flow pattern, it is convenient to introduce a conical
coordinate system  $\overrightarrow{r}\left( r, \theta\right) $, 
where r is the distance
measured from the cone vertex along the jet axis and
$ \theta$  is the angular displacement from the axis. The
directly observable jet region closest to the compact
object will be called the jet base. The location $ r_0 $,
electron density $ n_{e0}$, and temperature $T_{0} $ of this region
as well as the opening angle $ \Theta$\footnote{
In what follows, the temperature is in energy units and the
opening angle is in radians. In this case, since the latter is
small, we assume that $tan ~ \Theta\approx\Theta$.}  
 are the input parameters of the model.

Given the uniformity of the matter distribution in
the $ r=const$ layer, the law of change of the electron
density follows from the condition for the conservation
of mass flux along the jet:
\begin{equation}
	\label{nprof}
   n_{e}=n_{e0} \left( \frac{r}{r_{0}}\right) ^{-2}.
\end{equation}

	Using the derived density profile, we can write the
thermal balance equation as
\begin{equation}
   \frac{dT}{dr}=-2\left(\gamma-1 \right)  \frac{T}{r}-\frac{2n_{e} n_{i}}{3\left(n_{e}+n_{i}\right)} \frac{\Lambda \left( T \right)}{v},\label{cooling}
\end{equation}

where the first and second terms on the righthand
side correspond to the cooling through adiabatic
expansion and the losses through radiation,
respectively. Introducing dimensionless quantities
 $\eta$=T/T$_{0}$, $\xi$=r/r$_{0}$, X=n$_{i}$/n$_{e}$, $\Lambda_{23}\left(\eta\right)$=$\Lambda \left(\eta T_{0}\right)$/(10$^{-23}\frac{\text{erg} \ \text{cm}^{3}}{\text{s}}$) 
and assuming that $\gamma$=5/3, we obtain

\begin{equation}
	\label{unitlesscooling}
	\frac{d\eta}{d\xi}=-\frac{4}{3}\frac{\eta}{\xi}-\alpha \frac{\Lambda _{23} \left(\eta \right)}{\xi ^{2}},
\end{equation}	
\begin{equation}
	 \alpha = 10 ^{-23} \frac{\text{erg} ~ \text{cm}^{3}}{\text{s}} \cdot \frac{2}{3} \frac{n_{e0}r_{0}}{v T_{0}} \frac{X}{1+X}.
\end{equation}
Substituting the input parameters typical of the jet in
SS 433 and assuming that $X \approx0.91 $, we find
\begin{equation}
	\alpha \approx 0.1272 \left(\frac{n_{e0}}{10^{14} \text{cm}^{-3}}\right)\ \left(\frac{r_{0}}{10^{11} \text{cm}}\right)\ \left(\frac{T_{0}}{20 \text{keV}}\right)^{-1}\ \left(\frac{v}{0.26c}\right)^{-1}
\end{equation}

The limits of small and large $ \alpha$ correspond to adiabatic
and quasi-cylindrical jets, respectively (see below
and Section 4.3).\

The emissivity $\Lambda \left(T\right)$ in Eq. (\ref{cooling}) is calculated in
the model of a hot, optically thin plasma using the
low-density limit (APEC, \cite{smith01}) by assuming
solar elemental abundances (\citep{lodd03}).
A numerical solution of Eq. (\ref{unitlesscooling}) with the parameter
$\alpha$ and the initial condition $ \eta_{\alpha}\mid_{\xi=1}=1$ gives the
temperature profile along the jet $\eta_{\alpha}\left(\xi\right)$ or $T \left( r\right)=T_{0}\eta_{\alpha}\left(r/r_{0}\right)$.
 Since $\eta_{\alpha}\left(\xi\right)$ is a monotonic function of
$\xi$, there exists an inverse function $\xi_{\alpha}\left(\eta\right)$. 
This allows the boundary of the computational domain $\xi_{max}$ to
be determined, because thermal instabilities emerge
in the jet at $T_{min}\sim 0.1 $ keV \citep{kotani96}:
$\xi_{max}\left(\alpha\right)=\xi_{\alpha}\left(T_{min}/T_{0}\right)$.

The number of free parameters can be reduced by
fixing the total X-ray luminosity:
\begin{equation}
\label{gauge}
	L_{X}=n_{e0}^2 \ r_{0}^3 \ \Theta ^{2} I\left( \alpha \right),
\end{equation}
where
 \begin{equation}
	I \left(\alpha \right)=\pi X \int\limits_{1}^{\xi_{max}\left(\alpha\right)}
	\frac{\Lambda\left(T_{0}\eta\left(\xi ,\alpha\right)\right)}{\xi^{2}}d\xi 
\end{equation}

The cooling at low $ \alpha$ is determined by adiabatic
expansion, i.e., $\eta_{\alpha}\left(\xi\right)\approx\eta_{0}\left(\xi\right)=\xi^{-4/3}$ and $\xi_{max}=\left(T_{min}/T_{0}\right)^{-3/4} $ do not depend on $ \alpha$. Consequently,
$I\left(\alpha\right)\approx const$ and L$_{X}=const$ implies
\begin{equation}
\label{gaugeA}
	n_{e0}^2 \ r_{0}^3 \ \Theta ^{2} = const.
\end{equation}

At large $ \alpha$, in view of Eq. (\ref{unitlesscooling})
$\frac{\Lambda\left(T_{0}\eta\left(\xi ,\alpha\right)\right)}{\xi^{2}}d\xi  \propto \frac{d\eta}{\alpha}$ .
Consequently,  $I\left(\alpha\right)\propto 1/\alpha$ and L$_{X}=const$ implies
\begin{equation}
\label{gaugeB}
	n_{e0} \ r_{0}^2 \ \Theta ^{2} = const,
\end{equation}
which also corresponds to a constant mass loss rate in the jet $\dot{M}_{j}=const$.
%%%%%%%%%%%%%%%%%%%%%%%%%%%%%%%%%%%%
\section{RADIATIVE TRANSFER IN LINES} 
  It is convenient to investigate the radiative transfer
in a comoving frame of reference by getting rid of
the constant longitudinal velocity component. The
velocity field in such a frame is
\begin{equation}
\label{velfield}
	v_{\|}\left(\overrightarrow{r} \right)=0 , \ v_{\bot}\left(\overrightarrow{r} \right)=v_{j}*tan\left(\theta \right) ,\ v_{j}=0.26 c	
\end{equation}\	

In the introduced frame of reference, we will consider
a line photon with energy $E$ emitted on the jet
axis in a direction  $\overrightarrow{\Omega_{0}}$ perpendicular to the jet axis.
Since $ E, T\ll m_{e}c^{2}=511$ keV in the situation of interest
to us, the optical depth to the jet edge for scattering
by free electrons for such a photon is $\tau_{T}\left(\xi\right)=n_{e}\left(\xi\right)\sigma_{T}r_{0}\Theta\xi $,
 where $\sigma_{T}=6.65 \cdot 10^{-25}$ cm$^{2}$ is the
Thomson cross section. The calculation of the optical
depth to the jet edge for resonant scattering on
electron transitions in ions is slightly complicated by
the velocity field (\ref{velfield}). Therefore, we will initially
assume that $E=E_{0}$, where $E_{0}$ is the energy of the
corresponding transition. Given the Doppler shift in
the velocity field (\ref{velfield}) and the fact that the direction $\overrightarrow{\Omega_{0}}$
at each point is along the local velocity of the matter,
we will then obtain
\begin{equation}
\label{dtau}
d\tau_{res}\left(\theta ,\xi\right)=n_{i}\left(\xi\right)\sigma_{0}\ exp\left(-
\left(\frac{v_{\bot}\left(\theta\right)E_{0}}{\ \ c \ \ \  \Delta E_{D}}\right)^{2}\right)
r_{0}\xi d\theta ,
\end{equation}
where $\theta\in\left[0,\Theta\right]$ and, in view of the smallness of $\Theta$,
along with $\theta$, we set $tan\left(\theta\right)\approx\theta$. We also used the
following standard notation: $\sigma_{0}$ is the resonant scattering
cross section at the line center and $\triangle E_{D}$ is the
line width due to the thermal motion of ions (for more
details, see Section \ref{scattering}).

\ Denoting $x=\frac{\beta E_{0}\Theta}{\triangle E_{D}}$ ($\beta=v_{j}/c=0.26$), $y=x\frac{\theta}{\Theta}$,  $\tau_{res}\left(\xi\right)=n_{i}\left(\xi\right)\sigma_{0}r_{0}\Theta\xi $
 and integrating over $y$, we
will obtain the total optical depth to the jet edge for resonant scattering 
for a photon emitted on the axis
in the direction $\overrightarrow{\Omega_{0}}$:

\begin{equation}
\label{tauR}
\tau_{eff}\left(\xi\right)=\tau_{res}(\xi)\frac{\int\limits_{0}^{x} e^{-y^2}dy}{x} =\tau_{res}(\xi)\frac{\Phi\left(x\right)}{x},
\end{equation}
where $\Phi\left(x\right)=\frac{\sqrt{\pi}}{2} erf(x)$, $erf(x)$ is the standard error
function. \

\begin{figure}[h!]
\centering
\includegraphics[width=1.0\columnwidth]{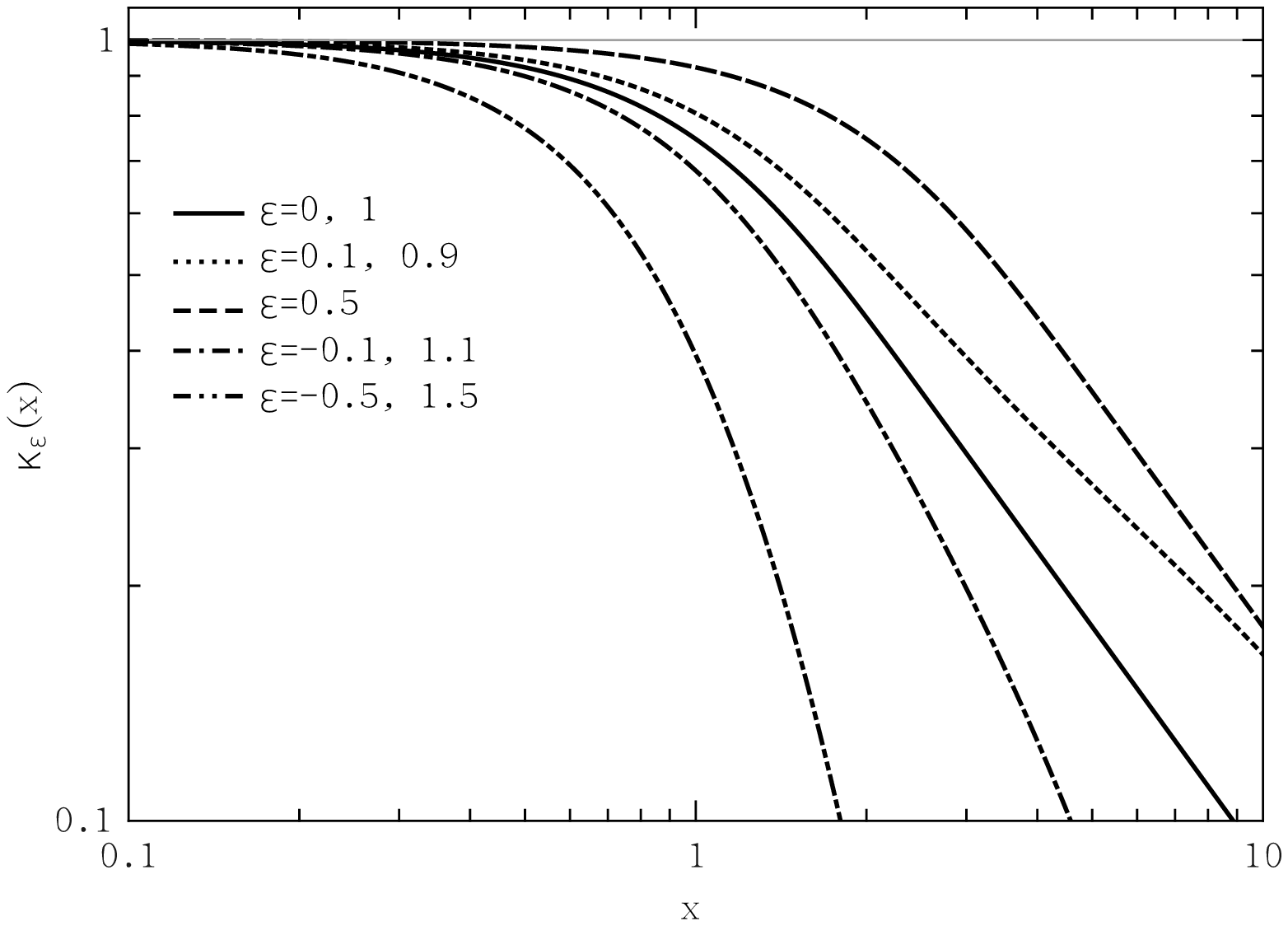}
\topcaption{General view of the function $\kappa_{\epsilon}\left(x\right)$, characterizing
the resonant scattering efficiency as a function of  $x=\frac{\beta E_{0}\Theta}{\triangle E_{D}}$ and the shift parameter $\epsilon=\frac{\Delta E_{0}}{\beta \Theta E_{0}}$ according to Eq. (\ref{tauRg})}
\label{kappa}
\end{figure}

Let us now consider a photon also emitted on the
jet axis in the direction $\overrightarrow{\Omega_{0}}$
 but with an initial energy
shift $\Delta E_{0}$ relative to the central value $E_{0}$, i.e., 
$E=E_{0}+\Delta E_{0}$. We see that for such a photon

\begin{equation}
\label{dtaue}
d\tau_{res}\left(\theta ,\xi\right)=n_{i}\left(\xi\right)\sigma_{0} e^{-\left(\frac{\Delta E_{0}-\beta E_{0}\theta}{\Delta E_{D}}\right)^{2}}r_{0}\xi d\theta ,
\end{equation}
and the total optical depth to the jet edge
\begin{equation}
\label{tauRe}
\begin{split}
 {\tau_{eff}\left(\xi ,\Delta E_{0}\right)=\tau_{res}(\xi)\frac{\int\limits_{-\epsilon x}^{(1-\epsilon)x} e^{-y^2}dy}{x} =}\\
={\tau_{res}(\xi)\frac{\Phi\left((1-\epsilon)x\right)+\Phi\left(\epsilon x\right)}{x}},
\end{split}
\end{equation}

where $\epsilon=\frac{\Delta E_{0}}{\beta \Theta E_{0}}$
is the initial shift parameter (in addition,
we used the oddness of erf(x)). Thus, the
effective optical depth for resonant scattering can
generally be written using the introduced notation as
\begin{equation}
\label{tauRg}
\tau_{eff}\left(\xi,\Delta E_{0}\right)=\kappa_{\epsilon}(x)\tau_{res}(\xi).
\end{equation}
The function $\kappa_{\epsilon}\left(x\right)=\frac{\Phi\left((1-\epsilon)x\right)+\Phi\left(\epsilon x\right)}{x}$ (see Fig. \ref{kappa}) is such that: \\
1) $\kappa_{\epsilon}\left(x\right)<1$ for any x and $\epsilon$;\\ 
2) $\kappa_{\epsilon}\left(x\right)$  is symmetric relative to $\epsilon=0.5$, i.e. $\kappa_{0.5-\delta\epsilon}\left(x\right)=\kappa_{0.5+\delta\epsilon}
\left(x\right)$ for any x and $\delta\epsilon$;\\
 3) for any x and $\epsilon\neq0.5$ $\kappa_{\epsilon}\left(x\right)< \kappa_{0.5}\left(x\right)$; \\
 4) at large x, $\kappa_{0}\left(x\right)\sim 1/x$; 
 
 \

\begin{figure}[h!]
\centering
\includegraphics[width=0.9\columnwidth]{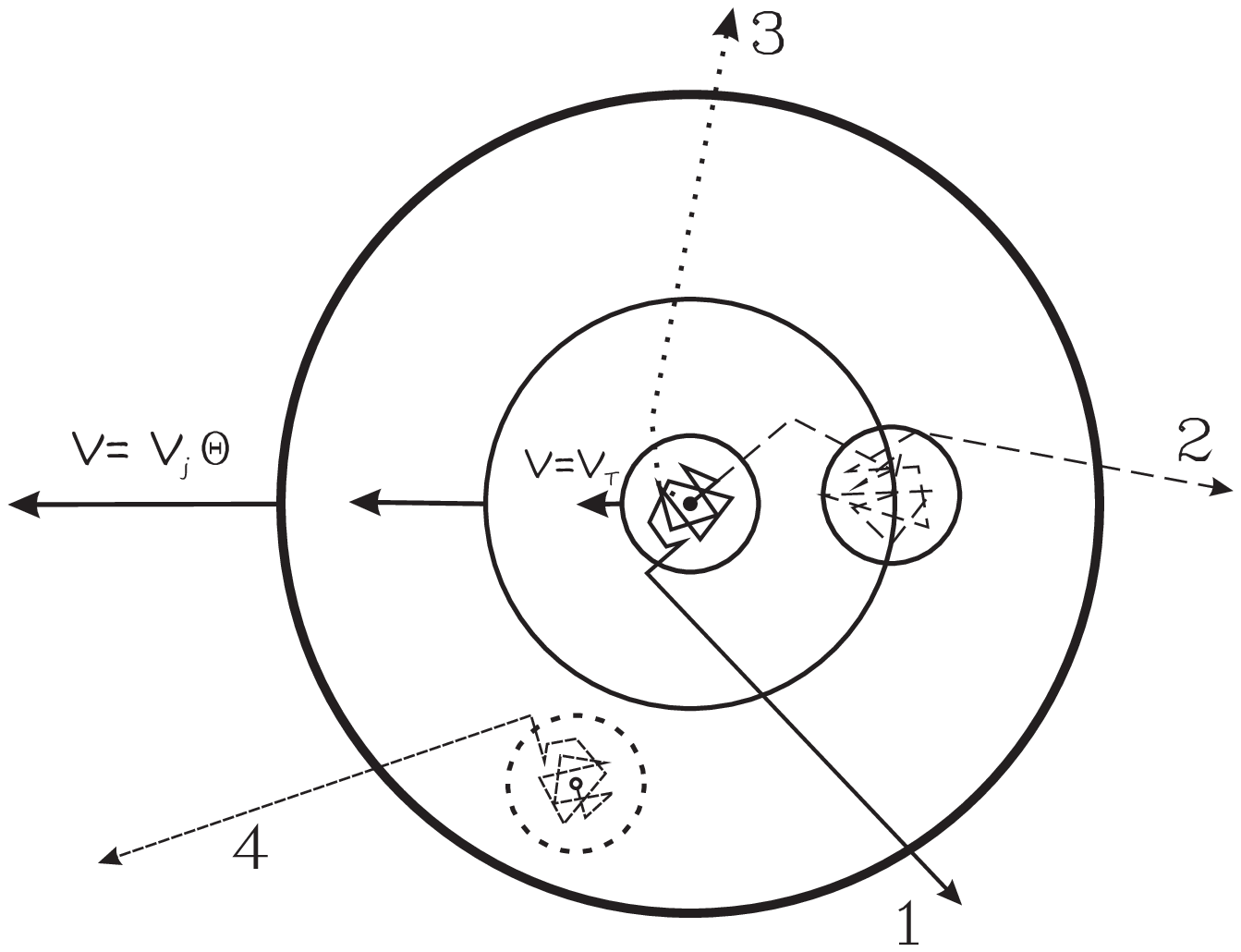}
\topcaption{Resonant scattering of photons in a layer perpendicular
to the jet axis for $x=\frac{\beta E_{0}\Theta}{\triangle E_{D}}=\frac{V_{j}\Theta}{V_{T}}\sim 5$. 
1,2,3 -- the pattern of the trajectory for photons emitted on the
axis with $E=E_{0},
E>E_{0},E<E_{0}$ respectively; 4 --
the pattern of the trajectory for a photon emitted not
on the axis with $E=E_{0}$ } 
\label{sketch}
\end{figure}

Thus, it can be concluded that not the entire optical
depth $\tau_{res}$ but only some part of it $\tau_{eff}$ is involved in
the resonant scattering of the photon under consideration;
the smaller the ratio of the thermal ion velocity
to the bulk one specified by the velocity field (\ref{velfield}), the
smaller this part. In view of the asymptotics defined by
property 4), this effect for a photon at the line center
($\epsilon=0$) is interpreted in such a way that the scattering
at large $ x$ occurs only inside the cylinder at the
boundary of which the bulk ion velocity is equal to the
thermal one. This means that the estimate of $\tau_{eff}\left(\xi,0\right)$
remains valid not only for the photons emitted on the
jet axis but also at any other point, except for the
surface layer of thickness $r_{0}\Theta\xi/x$. In addition, in view
of properties (2) and (3), we can assume the profiles to
be asymmetric relative to the central value due to the
larger effective optical depth for high-energy photons
($E>E_{0}, \epsilon>0$) (as an illustration of these properties,
see Fig. \ref{sketch}).

\

Let now the distribution of sources along the jet
be defined by some function $\phi\left(\xi\right)$, i.e., the relative
contribution of the layer $d\xi$ to the total photon luminosity
in the line is $\phi\left(\xi\right)d\xi$, and	
$\int\limits_{1}^{\xi_{max}}\phi\left(\xi\right)d\xi=1$.
The weighted mean optical depths
\begin{equation}
\label{tauTmean}
\widehat{\tau}_{T}=\int\limits_{1}^{\xi_{max}} \tau_{T}\left(\xi\right)\phi\left(\xi\right)d\xi ,
\end{equation}
\begin{equation}
\label{tauRmean}
\widehat{\tau}_{eff}=\int\limits_{1}^{\xi_{max}} \tau_{eff}\left(\xi,\Delta E_{0}\right)\phi\left(\xi\right)d\xi
\end{equation}
 then characterize the influence of scattering on the
integrated line emission.\

To estimate the probability of the scattering by
a free electron, we can use the critical optical depth
$\tau_{cr}=\frac{1}{2\ ln\left(\widehat{\tau}_{eff}/\widehat{\tau}_{T}\right)}$
following \cite{pozetal83}.
At $\widehat{\tau}_{T}<\tau_{cr}<1$, the jet may be considered optically
thin for scattering by electrons even for resonant
photons. At $\tau_{cr}<\widehat{\tau}_{T}<1$, almost all of the photons
will be scattered by the electron only once and, having
left the line, form broad wings.\

Undoubtedly, these estimates are qualitative in
nature due to the dependence of $\tau_{eff}$ on photon energy,
the distribution of sources inside the layer $d\xi$, the
velocity field (\ref{velfield}), and the line overlap in the case
of doublets and triplets. For a detailed analysis of
the line formation, we have to resort to numerical
simulations.\

Since this work is aimed not at a theoretical study
of the radiative transfer in lines but rather at its application
for a specific physical object, we will restrict
ourselves to the presented analysis, because the estimates
obtained in its context can give a clear interpretation
of the results obtained in our simulations
(see the ``Results'' Section).
%%%%%%%%%%%%%%%%%%%%%%%%%%%%%%%%%%%%%%%%%%%%%%
\section{SPECTRUM SIMULATION}
%%%%%%%%%%%%%%%%%%%%%%%%%%%%%%%%%%%%%%%%%%%%%%
\subsection{The Scheme of Computation}

The approach used was developed by \cite{pozetal83}
 and, subsequently, was successfully applied
to investigate the radiative transfer in galaxy
clusters \citep{saz02,zhur10}.
\subsubsection{Emission}

We will be interested in the
radiation from a jet in the line corresponding to the
transition from the upper level (UL) to the lower
level (LL) in an n-fold ionized atom of element Z
(for example, for the resonance K$_{\alpha}$ lines of heliumlike
atoms n = Z--2, UL= 1s2p($^{1}P_{1}$) and LL =
1s$^{2}$($^{1}S_{0}$).\

The dependences $n_{e}\left(\overrightarrow{r}\right)$ and 
$T\left(\overrightarrow{r}\right)$ found above
allow the local ionization balance to be computed in
the coronal approximation using AtomDB 2.0.1\footnote{http://atomdb.org/; }.
 As a result, in each $\xi=const$ layer of thickness d$\xi$ ($d\xi\ll\Theta\xi$) we will obtain the density of ions of a given type
$n_{i}\left(\xi\right)$ and the plasma emissivity $J\left(T\right)$ in the line of
interest to us at temperature $T=T_{0}\eta_{\alpha}\left(\xi\right)$. The line
photon production rate in such a layer is then defined
as
\begin{equation}
dL\left(\xi\right)=\pi  n_{e0}^{2}  r_{0}^{3} \Theta^{2}\ \ \frac{J\left(T_{0}\eta_{\alpha}\left(\xi\right)\right)}{\xi^{2}}d\xi.
\end{equation}
Thus, $ \phi\left(\xi\right)=\frac{1}{L_{0}}\frac{dL}{d\xi}$, where the total luminosity
\begin{equation}
L_{0}=\pi n_{e0}^{2}  r_{0}^{3} \Theta^{2} \int\limits_{1}^{\xi_{max}\left(\alpha\right)}
	\frac{J\left(T_{0}\eta_{\alpha}\left(\xi \right)\right)}{\xi^{2}}d\xi .
\end{equation} 
\	Similarly, for the continuum radiation 
\begin{equation}
L_{c} \left(E\right)=\pi n_{e0}^{2}  r_{0}^{3} \Theta^{2} \int\limits_{1}^{\xi_{max}\left(\alpha\right)}
	\frac{J_{c}\left(E,T_{0}\eta_{\alpha}\left(\xi \right)\right)}{\xi^{2}}d\xi ,
\end{equation} 
 
\ where $J_{c}\left(E,T\right)$ is the plasma emissivity in the continuum
per unit energy interval with center $  E$ at
temperature $ T $ computed using the NoLine model of
AtomDB 2.0.1 (kindly provided by Adam Foster \footnote{A description of the 
analogous model for APEC v1.3.1 can be found at http://cxc.harvard.edu/twiki/bin/view/SnrE0102/NoLine}). 
The equivalent width of a line with a transition energy $E_{0}$
is then defined as
\begin{equation}
\label{ew}
EW=L_{0}/L_{c}\left(E_{0}\right).
\end{equation} 

\ However, it is worth noting that the equivalent
width defined in this way has no direct bearing on
the observed quantities, because both jets contribute
to the actual continuum, given the Doppler shift and
relativistic collimation of the radiation. Nevertheless,
assuming the jets to be identical, determining the
observed equivalent width from the line equivalent
width that we use for an arbitrary precession phase
does not seem problematic.
\subsubsection{Scattering}
\label{scattering} 
	The reciprocal of the mean free
path (scattering coefficient) serves as a local characteristic
defining the photon scattering probability:
\begin{equation}
\Sigma_{x}=n_{x}\sigma_{x},
\end{equation}

where $n_{x}$  is the number density of scattering centers
and $\sigma_{x}$ is the cross section for the corresponding
scattering. For example, for scattering by electrons,
$\Sigma_{e}=n_{e}\sigma_{T} $, \ $\sigma_{T}=6.65 \cdot 10^{-25}$ cm$^{2}$
 is the Thomson cross section. For resonant scattering
by ions $\Sigma_{i}=n_{i}\sigma_{res} $,\ $\sigma_{res}=\sigma_{0} \ exp\left[- \left(\frac{\triangle E}{\triangle E_{D}}\right)^2\right] $,\ $\sigma_{0}=\frac{\sqrt{\pi} h r_{e}cf}{\triangle E_{D}}$,\ $\triangle E_{D}=E_{0}\left[\frac{2T}{Am_{p}c^{2}}\right]^{1/2}$,\ $\triangle E=E-E_{0}\left(1+\frac{v_{\perp}}{c}cos \psi\right)$,  where $r_{e}$ is the classical
electron radius, $m_{p}$ is the proton mass, $  A$ is 
the atomic weight of the ion, $  E$ is the photon energy, and
$\psi$ is the angle between the photon direction and the
local gas velocity determined by field (\ref{velfield}). The energy
$E_{0}$ and oscillator strength $  f$ of the atomic transition
can be found using AtomDB 2.0.1. It is worth noting
that although the oscillator strength $  f$ is a basically
positive quantity, in what follows (e.g., in Table 1), we
take $ f=0 $ for transitions with $f<0.0001  $ to simplify
our calculations.

Whereas $\Sigma_{e}$ depends only on the local electron
density, $\Sigma_{i} $ is a function of the local plasma characteristics
($n_{e}$ and $ T $), the photon energy and direction.

We performed detailed Monte Carlo simulations of
the emergent radiation spectrum in lines. We used
a scheme with a statistical ``weighting'' of photon
packets, along with the method of a constant total
cross section \citep{sob73}. The idea of this method
is to introduce, along with the real types of scattering,
some fictitious scattering, such that
\begin{equation}
\Sigma_{fict}=\Sigma_{0}-\left(\Sigma_{e}+\Sigma_{i}\right),
\end{equation} 

where the constant$\Sigma_{0}$ is greater than or equal to the
maximum possible $\Sigma_{real}=\Sigma_{e}+\Sigma_{i}$.
 As a result, the inhomogeneous (in the sense of local scattering characteristics)
medium is replaced by a homogeneous
one with a small mean free path $ \lambda=\Sigma_{0}^{-1}$. In each
scattering, the realization probability of a particular
type is defined as $ p_{x}=\Sigma_{x}/\Sigma_{0}$. In the case of fictitious
scattering, both direction and energy of the photon
package remain unchanged. The resonant scattering
by ions is represented as a combination of the dipole
and isotropic components. The weight of the dipole
component  $w_{d}$ is determined by the total angular
momentum of the lower level  $j_{LL}$ and by the difference
$\triangle j= j_{UL}-j_{LL}$ \citep{ham47}.\

The described scheme can be extended to the simulation
of unresolvable multiplets (the photons fall
into the total spectrum ``weighted'' proportionally to
the parent line intensity, while the scattering on each
multiplet component is considered as a separate type
of scattering). The addition of a broad continuum
component and the calculation of its scattering are
also possible.

\subsection{The Set of Lines}

	To form the set of simulated lines, the best resolution
Chandra HETGS spectra \citep{marshetal02,nametal03,lopezetal06} should
be used. The brightest observed lines subject to
strong scattering effects correspond to the $K_{\alpha}$ transitions
in helium-like ions and to the $Ly_{\alpha}$ transitions
in hydrogen-like ions with significant oscillator
strengths whose parameters are listed in Table \ref{linelist}. In
this case, the $K_{\alpha}$ transitions in helium-like ions have
a triplet structure (the resonance (w), intercombination
(x + y), and forbidden (z) lines). Bearing in
mind the importance of these lines for diagnosing
the jet plasma parameters, we performed detailed
simulations of the components by taking into account
the possible interaction via the optical depths and the
influence of collisions on the upper level population
of the forbidden and intercombination transitions at
$n_{e}$ greater than some $n_{e,crit}$ \citep{PD10}. In
addition, in some cases, the influence of satellites on
the intensity ratio of these lines should be taken into
account (see the ``Results'' Section).
	  
%%%%%%%%%%%%%%%%%%%%%%%%%%%%%%%%%%%%%%%%%%	 
\begin{table*}
\centering
\caption{The set of simulated lines. The asterisk  ($^{\ast}$) marks the lines for which we do not present our simulation results
but which are used in constructing the total broadband spectrum (see the Conclusions). The expression $ f=0 $ should be
understood in the sense that  $f < 0.0001 $ (see also the text)
\centerline{}}
\begin{tabular}{ccccc}\hline\hline
{Spectroscopic} &\multirow{2}*{Transition}&\multirow{2}*{$E_{0}$, keV}&\multirow{2}*{f}&\multirow{2}*{$w_{d}$}\\
{symbol} &{}&{}&{}&{}				
\\[2 mm]\hline \\
{Fe XXV}&{$K_{\beta}:1s^{2}-1s3p$ ($^{1}P_{1}$)}&{7.881}&{0.14}&{1}\\[2mm]
% Ni Ka complex
{Ni XXVII}&{$K_{\alpha}:1s^{2}-1s2p$ ($^{1}P_{1}$)(w)}&{7.806}&{0.72}&{1}\\
{Ni XXVII}&{$K_{\alpha}:1s^{2}-1s2p$ ($^{3}P_{2}$)(x)}&{7.799}&{0}&{-}\\
{Ni XXVII}&{$K_{\alpha}:1s^{2}-1s2p$ ($^{3}P_{1}$)(y)}&{7.766}&{0.07}&{1}\\
{Ni XXVII}&{$K_{\alpha}:1s^{2}-1s2p$ ($^{3}S_{1}$)(z)}&{7.744}&{0}&{-}\\[2mm]
% Fe Lya complex
{Fe XXVI}&{$Ly_{\alpha}:1s-2p$ ($^{1}P_{3/2}$)}&{6.973}&{0.27}&{0.5}\\
{Fe XXVI}&{$Ly_{\alpha}:1s-2p$ ($^{1}P_{1/2}$)}&{6.952}&{0.14}&{0}\\[2mm]
% Fe Ka complex
{Fe XXV}&{$K_{\alpha}:1s^{2}-1s2p$ ($^{1}P_{1}$)(w)}&{6.700}&{0.78}&{1}\\
{Fe XXV}&{$K_{\alpha}:1s^{2}-1s2p$ ($^{3}P_{2}$)(x)}&{6.682}&{0}&{-}\\
{Fe XXV}&{$K_{\alpha}:1s^{2}-1s2p$ ($^{3}P_{1}$)(y)}&{6.667}&{0.07}&{1}\\
{Fe XXV}&{$K_{\alpha}:1s^{2}-1s2p$ ($^{3}S_{1}$)(z)}&{6.636}&{0}&{-}\\[2mm]
% Ca & Ar 
{Ca XIX \protect\footnote{}\saveFN\oldfoot\par}&{$K_{\alpha}:1s^{2}-1s2p$ ($^{1}P_{1}$)}&{3.902}&{0.77}&{1}\\
{Ar XVII \protect\useFN\oldfoot}&{$K_{\alpha}:1s^{2}-1s2p$ ($^{1}P_{1}$)}&{3.133}&{0.77}&{1}\\[2mm]
% S Lya complex
{S XVI \protect\useFN\oldfoot}&{$Ly_{\alpha}:1s-2p$ ($^{1}P_{3/2}$)}&{2.623}&{0.27}&{0.5}\\
{S XVI \protect\useFN\oldfoot}&{$Ly_{\alpha}:1s-2p$ ($^{1}P_{1/2}$)}&{2.620}&{0.14}&{0}\\[2mm]
% S Ka complex
{S XV}&{$K_{\alpha}:1s^{2}-1s2p$ $^{1}P_{1}$)(w)}&{2.461}&{0.76}&{1}\\
{S XV}&{$K_{\alpha}:1s^{2}-1s2p$ ($^{3}P_{2}$)(x)}&{2.449}&{0}&{-}\\
{S XV}&{$K_{\alpha}:1s^{2}-1s2p$ ($^{3}P_{1}$)(y)}&{2.447}&{0.07}&{1}\\
{S XV}&{$K_{\alpha}:1s^{2}-1s2p$ ($^{3}S_{1}$)(z)}&{2.430}&{0}&{-}\\[2mm]
% Si Lya complex
{Si XIV \protect\useFN\oldfoot}&{$Ly_{\alpha}:1s-2p$ ($^{1}P_{3/2}$)}&{2.006}&{0.27}&{0.5}\\
{Si XIV \protect\useFN\oldfoot}&{$Ly_{\alpha}:1s-2p$ 
($^{1}P_{1/2}$)}&{2.004}&{0.14}&{0}\\[2mm]
% Si Ka complex
{Si XIII}&{$K_{\alpha}:1s^{2}-1s2p$ ($^{1}P_{1}$)(w)}&{1.865}&{0.75}&{1}\\
{Si XIII}&{$K_{\alpha}:1s^{2}-1s2p$ ($^{3}P_{2}$)(x)}&{1.855}&{0}&{-}\\
{Si XIII}&{$K_{\alpha}:1s^{2}-1s2p$ ($^{3}P_{1}$)(y)}&{1.854}&{0.07}&{1}\\
{Si XIII}&{$K_{\alpha}:1s^{2}-1s2p$ ($^{3}S_{1}$)(z)}&{1.839}&{0}&{-}\\[2mm]
\hline
\end{tabular}
\label{linelist}
\end{table*}
%\newpage
%%%%%%%%%%%%%%%%%%%%%%%%%%%%%%%%%%%%%%%%%%
\subsection{Input parameters}

\ 	There is a considerable uncertainty in the physical
parameters of the jets in SS 433. Based on
Chandra HETGS observations at the phase of the
greatest disk opening toward the observer, \cite{marshetal02}
obtained $r_{0}\simeq 2 \times 10^{10}$ cm,  $n_{e0}\simeq 2 \times 10^{15}$ cm$^{-3}$, $T_{0}\simeq 13$ keV, and $\Theta \simeq 0.01$ rad for the
approaching jet. Analysis of the XMM-Newton X-ray
spectra also at the phase of the greatest opening
\citep{medved10} gives $r_{0}\simeq 2 \times 10^{11}$ cm
and $T_{0}\simeq 17$ keV for the approaching jet. In contrast,
studies based on the eclipse of the jets by the optical
companion yield $r_{0}\sim 1 \times 10^{12}$ cm and $T_{0}\sim 30$ keV
\citep{kateetal06}.

\	Therefore, two classes of models satisfying condition
(\ref{gauge}) for the conservation of total X-ray luminosity
were formed -- for the pair of jets $L_{x}\sim 10^{36} erg/s $
\citep{medved10} in the range from 0.1 to
 50 keV either in form (\ref{gaugeA}) (the class of quasi-adiabatic
models) or in form (\ref{gaugeB}) (the class of quasi-cylindrical
models), depending on the parameter $\alpha$. Since the
available observations can reliably set only an upper
limit on the opening angle (see Section 6),
we allowed $\Theta$ to change in a very wide range. For
this reason, precisely $\Theta$ served as a class-forming
parameter (in the sense of discrimination by $\alpha$). As
a result, the simulations were performed on a wide
grid of input parameters given in Table \ref{GaugeTableX} and encompassing
the most probable values. The quasiadiabatic
models correspond to $\Theta=0.01$ rad and
$\Theta=0.02$ rad, while the quasi-cylindrical ones 
correspond to $\Theta=0.003$ rad and $\Theta=0.0001$ rad. We
considered $T_{0}$=20 keV and  $T_{0}$=30 keV as possible
plasma temperatures at the jet base. The abundances
of heavy elements were assumed to be solar \citep{lodd03}.\

It is worth noting that whereas in the case of
quasi-adiabatic models $r_{0}$ roughly coincides with the
distance from the jet base to the compact object,
in the case of quasi-cylindrical models $r_{0}$ has the
meaning of only a geometrical parameter of the flow. Therefore,
using the combination $r_{0}\Theta$, which means the
transverse size of the jet near its base, appears more
preferable for the description of a particular model in
the general case.
%%%%%%%%%%%%%%%%%%%%%%%%%%%%%%%%%%%%%%%%
%%%%							GAUGE TABLES                                        %%%%	
%%%%%%%%%%%%%%%%%%%%%%%%%%%%%%%%%%%%%%%%

\begin{figure}[h!]
\centering
\includegraphics[width=1.0\columnwidth]{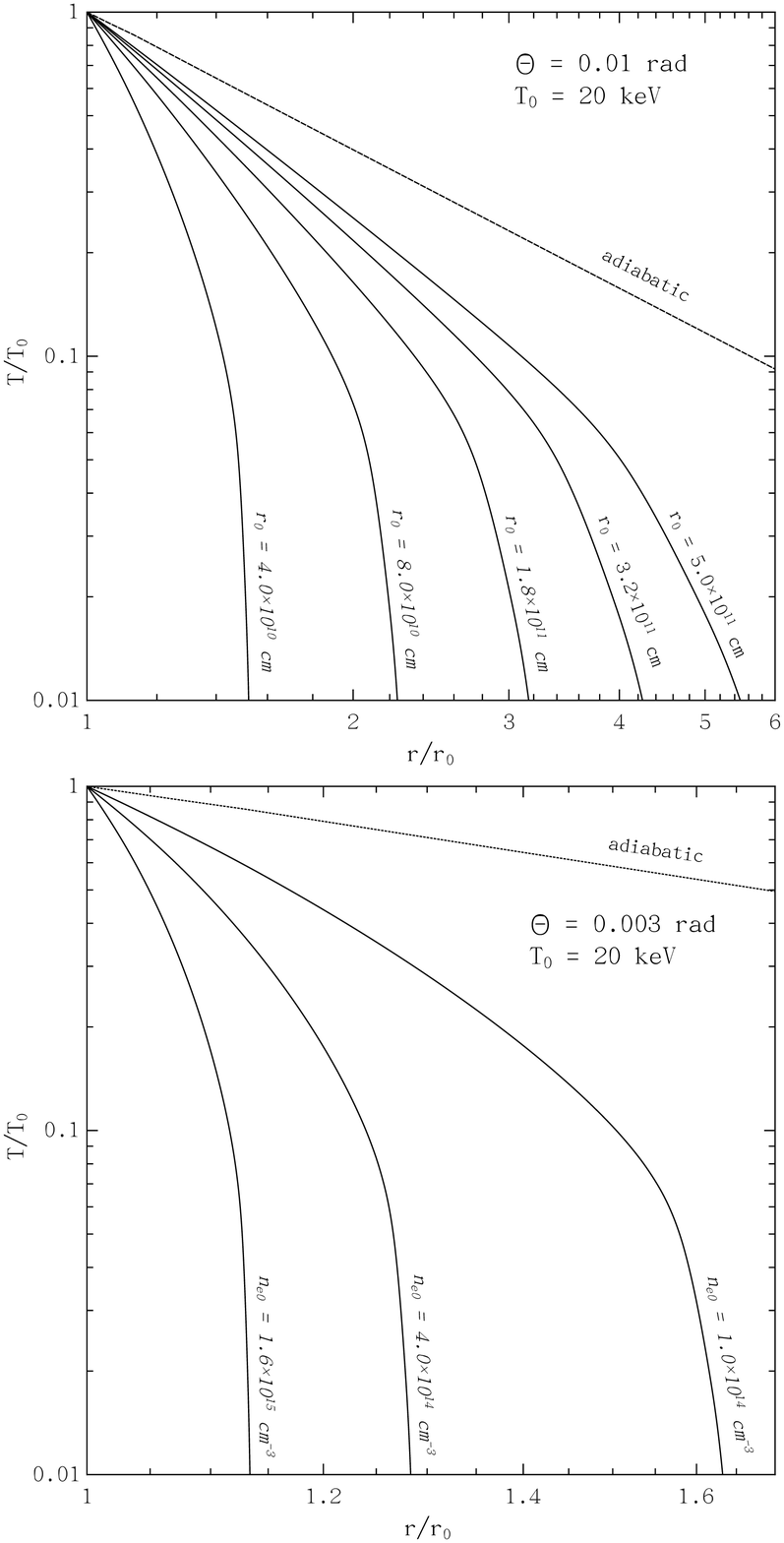}
\topcaption{Temperature profiles along the jet for various input
parameters from Table \ref{GaugeTableX}}
\label{tempprof}
\end{figure}

\begin{figure*}[h!]
%\begin{flushleft}
\epsfxsize=1.08\columnwidth \epsfbox{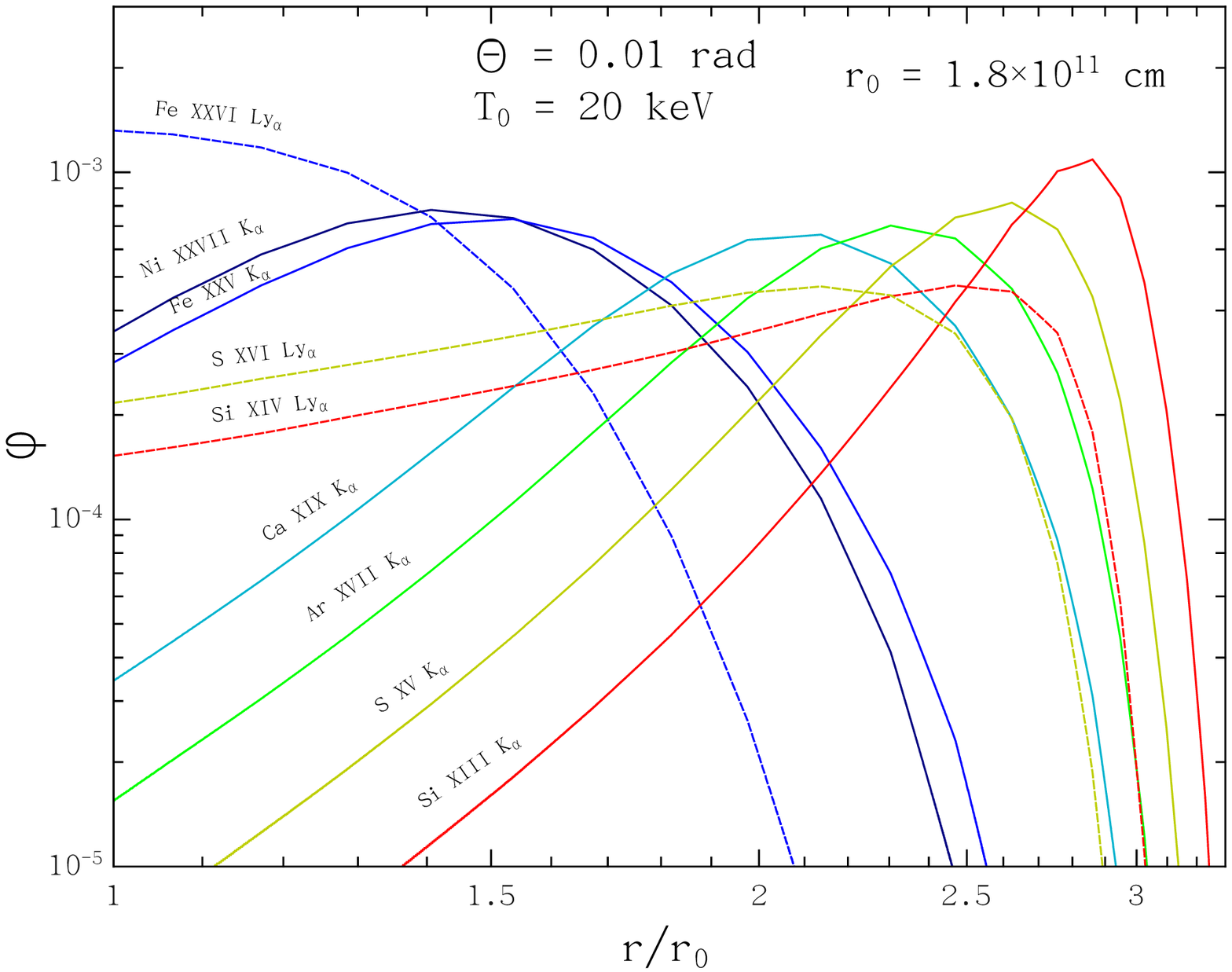}
%\end{flushleft}
%\centering
\epsfxsize=1.2\columnwidth \epsfbox{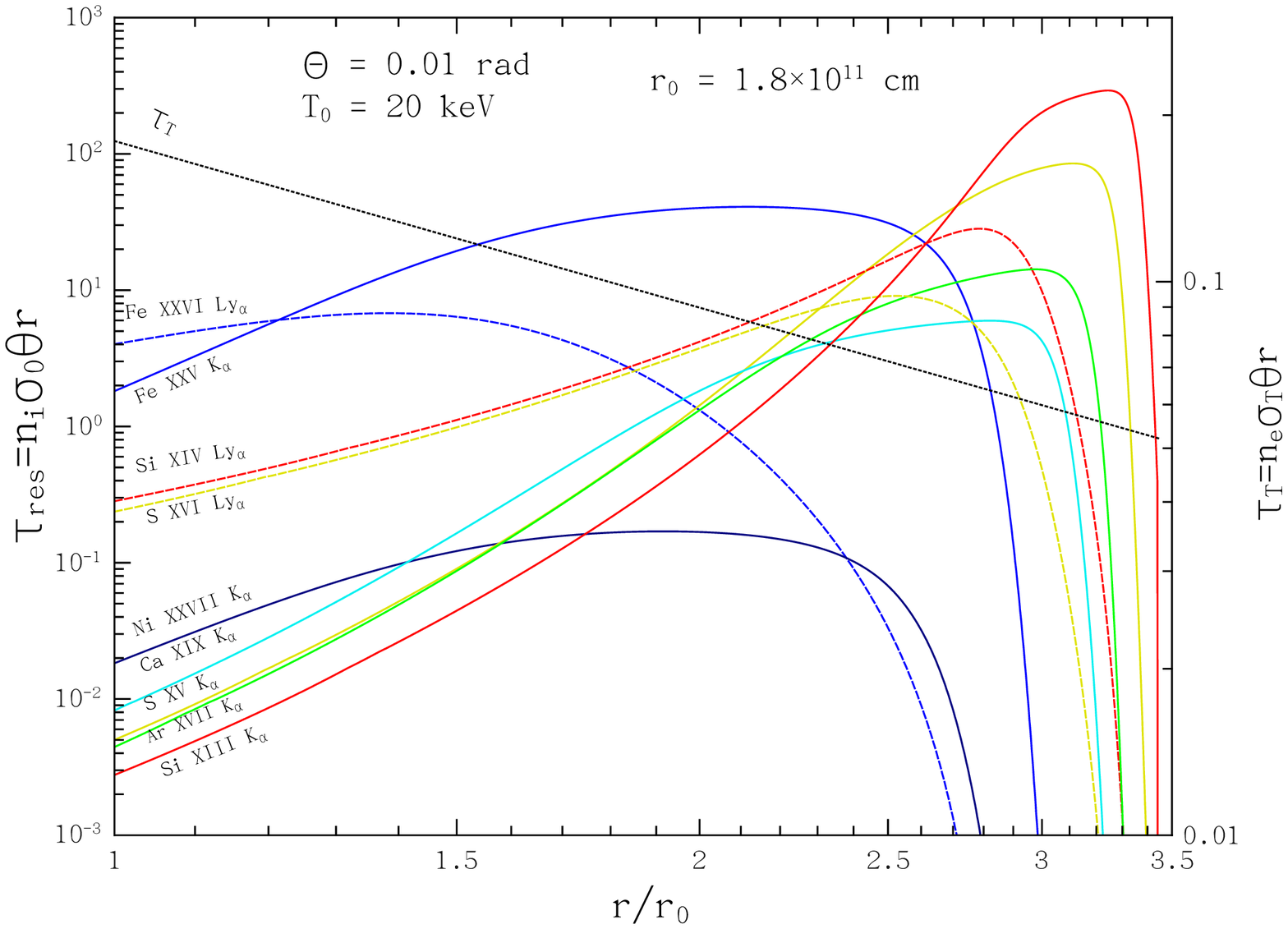}
\caption{\textbf{Left} -- Contribution from various jet regions to the total luminosity $ \phi\left(\xi\right)=\frac{1}{L_{0}}\frac{dL}{d\xi}$ 
in some of the lines from Table \ref{linelist}. \textbf{Right} --
Profiles of the transverse optical depth of the jet for scattering by electrons and resonant scattering in some of the lines from
Table \ref{linelist}.}

\label{fluxdepthprof}
\end{figure*}

\begin{table*}[hb]
\topcaption{The grid of input parameters and the corresponding values of $\alpha$  at $T_{0}=20$ keV as well as the luminosities
($ L_{n}=L/10^{n} $) and equivalent widths (without scattering) of the lines from Table 1 for which detailed simulation results
are presented (see the ``Results'' Section). The use of the quasi-cylindrical calibration (\ref{gaugeB}) when passing from $r_{0}=8\times 10^{10}$ cm to $r_{0}=4\times 10^{10}$cm for $\Theta=0.01$ and
$\Theta=0.02$ rad is marked ($^{\ast}$) , because $\alpha\sim 1$ in this case}
\centering
\begin{tabular}{cc|c|cccccccccccc}\hline\hline
{$r_{0}\Theta $ }&{$n_{e0}$  } &\multirow{3}*{$\alpha$}
&\multicolumn{2}{c}{Fe XXV K$_{\alpha}$}
&\multicolumn{2}{c}{Fe XXVI Ly$_{\alpha}$}
&\multicolumn{2}{c}{Ni XXVII K$_{\alpha}$}
&\multicolumn{2}{c}{Fe XXV K$_{\beta}$}
&\multicolumn{2}{c}{S XV K$_{\alpha}$}
&\multicolumn{2}{c}{Si XIII K$_{\alpha}$}\\
{$10^{9} $}&{$ 10^{14}$}&{}
&{L$_{33}  $,}&{EW}
&{L$_{33}  $}&{EW }
&{L$_{32}  $}&{EW}
&{L$_{32}  $}&{EW }
&{L$_{33}  $}&{EW}
&{L$_{33}  $}&{EW}\\
{cm}&{cm$^{-3}$}&{}&
{erg/s}&{eV}&{erg/s}&{eV}&
{ erg/s}&{eV}&{erg/s}&{eV}&
{ erg/s}&{eV}&{erg/s}&{eV}
%$ 10^{14}$ 
%$10^{33}$
\\[2mm]\hline\hline
\multicolumn{15}{c}{Quasi-adiabatic models}\\
\hline
%%%%%%%%%%%%%%%%%
\multicolumn{15}{c}{$ \Theta=0.02 $ rad }\\
\hline
{10.0}&{0.16}&{0.10}&{8.7}&{227}&5.00 &164& 5.59 & 21& 9.11& 34& 0.99& 13.3& 1.48& 15.4\\
{6.4}&{0.31}&{0.13}&{8.4}&{230}& 4.73& 163 & 5.34& 21 &8.74 & 35& 0.91& 13.7& 1.44& 15.8\\
{3.6}&{0.75}&{0.17}&{8.4}&{234}& 4.63& 163& 5.34& 22 & 8.7& 35& 0.85 & 14.2& 1.44& 15.9\\
{1.6}&{2.5}&{0.25}&{7.4}&{239}& 4.00 & 162& 4.66 & 22&7.7& 36& 0.65& 14.3& 1.23& 15.4\\
{0.8}&{10.0\protect\footnotemark
\footnotetext{}\saveFN\oldfootx}&{0.51}&{11.1}&{242}& 5.88& 162& 6.91& 22 & 11.47& 37& 0.98& 13.6& 1.61& 13.7\\
\hline
%%%%%%%%%%%%%%%%%
\multicolumn{15}{c}{$ \Theta=0.01 $ rad }\\
\hline
{5.0}&{0.32}&{0.20}&{7.8}&{238}& 4.29& 163 & 4.96& 22 & 8.17& 36& 1.04& 14.3& 1.34& 15.7 \\
{3.2}&{0.62}&{0.25}&{7.3}&{239}& 3.94& 162& 4.58& 22 & 7.57 & 36 & 1.03& 14.3& 1.21& 15.4\\
{1.8}&{1.5}&{0.34}&{6.9}&{241}& 3.68& 162&4.3& 22 & 7.13 & 36 & 1.05 & 14.1& 1.09 & 14.8\\
{0.8}&{5.0}&{0.51}&{5.6}&{242}& 2.95 & 162& 3.46 & 22&5.73  & 37 & 0.93& 13.6& 0.80& 13.7 \\
{0.4}&{20.0\protect\useFN\oldfootx}&{1.02}&{7.0}&{240}& 3.76& 162&4.38 & 22& 7.24 & 36 & 1.30& 12.4& 0.89& 12.1
\\[2mm]\hline\hline
\multicolumn{15}{c}{Quasi-cylindrical models}\\
\hline
%%%%%%%%%%%%%%%%%
\multicolumn{15}{c}{$ \Theta=0.003 $ rad }\\
\hline
{2.0}&{1.0}&{8.69}&{9.0}&{241}& 4.84  & 162&5.62 & 22 & 9.3& 36& 0.99 & 12.6& 1.17& 12.4\\
{1.0}&{4.0}&{17.38}&{10.1}&{237} & 5.54& 163&6.4 & 22 & 10.34& 36& 1.04& 11.9& 1.22& 11.3\\
{0.5}&{16.0}&{34.77}&{10.8}&{234}& 6.03 & 163&6.89 & 22 & 11.31& 35& 1.08& 11.4& 1.24& 10.8\\
\hline
%%%%%%%%%%%%%%%%%
\multicolumn{15}{c}{$ \Theta=0.0001 $ rad }\\
\hline
{2.0}&{1.0}&{26.07}&{11.5}&{231}& 6.59 & 163&7.46 &  21 &12.22& 35& 1.12& 10.9& 1.28 & 10.3\\
{1.0}&{4.0}&{52.15}&{11.5}&{231}& 6.52& 163&7.35 &  21 &12.03& 35& 1.10& 10.8& 1.26& 10.3\\
{0.5}&{16.0}&{104.30}&{11.5}&{231}& 6.52& 163 &7.37 & 21 & 12.06& 35& 1.10& 10.8& 1.26& 10.4\\

\hline
\end{tabular}
\label{GaugeTableX}
\end{table*}

%%%%%%%%%%%%%%%%%%%%%%%%%%%%%%%%%%%%%%%%
\	To illustrate the physical picture that emerges
when considering the constructed grid of parameters
through the prism of the jet model from Section 2,
it is convenient to use the profiles of temperature
$T \left( \xi \right)$ (Fig. \ref{tempprof}), optical depth $\tau_{x}\left(\xi\right)=\Theta\xi r_{0}\Sigma_{x}$, and
flux $\phi_{i}\left(\xi\right)$ (Fig. \ref{fluxdepthprof}) in lines.
%%%%%%%%%%%%%%%%%%%%%%%%%%%%%%%%%%%%%%%%
%%%%							Profiles                                                      %%%%	
%%%%%%%%%%%%%%%%%%%%%%%%%%%%%%%%%%%%%%%%

%\clearpage
%\newpage
%%%%%%%%%%%%%%%%%%%%%%%%%%%%%%%%%%%%%%%%%%%%
%%%%                 			RESULTS                                                         %%%%
%%%%%%%%%%%%%%%%%%%%%%%%%%%%%%%%%%%%%%%%%%%%
\section{Results}
\label{results}
%%%%%%%%%%%%%%%%%%%%%%%%%%%%%%%%%%%%%%%%%%%%%%
\subsection{Lines}

\ \ \ 	To save space, we present the results of our
computations only for the most important (in our
view) lines of the approaching jet with $ T_{0}=20$ keV
at the phase of the greatest disk opening that corresponds
to an angle between the jet axis and the
observer's direction $\chi \approx 65^{\circ} $, i.e., a redshift $ z_{b}=-0.078 $
 ($z=\gamma\left(1-\beta cos\chi\right)-1,~  \gamma=1/\sqrt{1-\beta^{2}},\ \beta=v_{j}/c=0.26$). The results for the remaining
(marked by $  ^\ast$ in Table \ref{linelist}) lines are used only in
constructing the total broadband spectrum (see the
Conclusions). Analysis of our computations for other
precession phases and $ T_{0}=30 $ keV showed that the
general conclusions (see the Conclusions) remain
valid here as well, while most of the changes are
equivalent to some shift on the constructed grid of
parameters in the sense of scattered line characteristics.\

The main quantitative characteristic reflecting the
influence of scattering effects on the observed spectrum
is the fraction of the photons $\zeta$ that left the
line as a result of their scattering by a free electron
and that fell into the broad wings. Together with
the temperature $T_{e}$ providing the best agreement between
the shape of the broad wings and the single
Compton scattering kernel \citep{saz00}, the parameter $\zeta$ allows the contribution
of scattered radiation to the continuum to be completely
described (for more details, see Section 5.2).
Since it is probably impossible to find such a complete
and universal way of describing the shape of the
scattered line itself, one has to content oneself with
standard approximations when analyzing the derived
profiles. This approach is justified from the viewpoint
of observational characteristics of existing spectrometers
(Chandra HETGS), which do not allow the
fine structure of the line profiles to be investigated in
most cases. However, it is clearly insufficient from
the viewpoint of new-generation X-ray observatories
(primarily Astro-H). For this reason, the set of parameters,
except for the line attenuation coefficient
equal to ($1-\zeta$), is determined by the specific simplifying
model that is most appropriate for the line
in the resolution of a particular instrument (Chandra
HETGS (to be more precise, HEG--High Energy
Gratings) or Astro-H SXS).

 To analyze the spectra in the Chandra HEG
resolution, we used a model response function in the
form of a Gaussian with $FWHM_{Ch}\left(E_{z}\right)=29.9$ eV
$\left(\frac{E_{z}}{6.4keV}\right)^{2}$
(in accordance with the Chandra Gratings
Fact Sheet \protect\footnote{https://icxc.harvard.edu/rws 
/peer\_review/proc\_docs/ \\ /Fact\_Sheet\_gratings.pdf
}), where $E_z = E_z/(1 + z)  $. For
Astro-H SXS, we used a Gaussian with
$FWHM_{AH}\left(E_{z}\right)=5 eV=const$ (in accordance with
the Astro-H Quick Reference \footnote{http://astro-h.isas.jaxa.jp/doc/ahqr.pdf}).\
 
  Since all lines exhibit very similar patterns in various
regimes of radiative transfer, we will restrict ourselves
to a detailed description and interpretation of
the results only for the $K_{\alpha}$ triplet of helium-like iron,
providing the scattering features in a specific line,
where necessary.
%%%%%%%%%%%%%%%%%%%%%%%%%%%%%%%%%%%%%%%%%
\subsubsection{Fe XXV K$_{\alpha}$}	
\label{fexxv}

The helium-like iron triplet
is brightest among the observed lines (Table \ref{GaugeTableX}). Nonetheless,
 only the components corresponding to the
allowed transitions with $f>0$ (see Table \ref{linelist}) can have
a significant intrinsic optical depth for resonant scattering.
At the same time, the line broadening due
to the transverse velocity component at $\Theta\sim 0.01$ rad
is enough to overlap the resonance and intercombination
lines. This makes the interaction between
the components via the optical depths possible and,
consequently, indicates that all components should
be taken into account simultaneously when the local
characteristics of the scattering medium are calculated.\

The computation performed with the Cloudy code
(version 08.00 described by \cite{ferland1998}) in
the temperature range of interest to us showed that for
helium-like iron $n_{e,crit}\sim 10^{17} $ cm$^{-3}$ is the density at
which the collisional excitations from $^{3}S_{1}$ (the upper
level of the forbidden component) to $^{3}P_{0,1,2}$ (the upper
levels of the intercombination components) begin to
dominate over the radiative transition from $^{3}S_{1}$ to
the ground level. Thus, on the specified grid of parameters
$n_{0}\ll n_{e,crit}$, i.e., the intensity redistribution
of the forbidden and intercombination lines for iron
may be neglected. Note that this conclusion is also
valid for the nickel triplet, because 
$n_{e,crit}=\frac{ A_{fg}}{C_{fi}}\propto Z^{13} $
 \citep{Mewe78a}, where $A_{fg}$ is the rate
of radiative transitions to the ground level and $C_{fi}$ is
the rate of collisional excitations to the upper levels of
the intercombination lines.\

\begin{figure*}[hb]
{\centering 
\smfigure{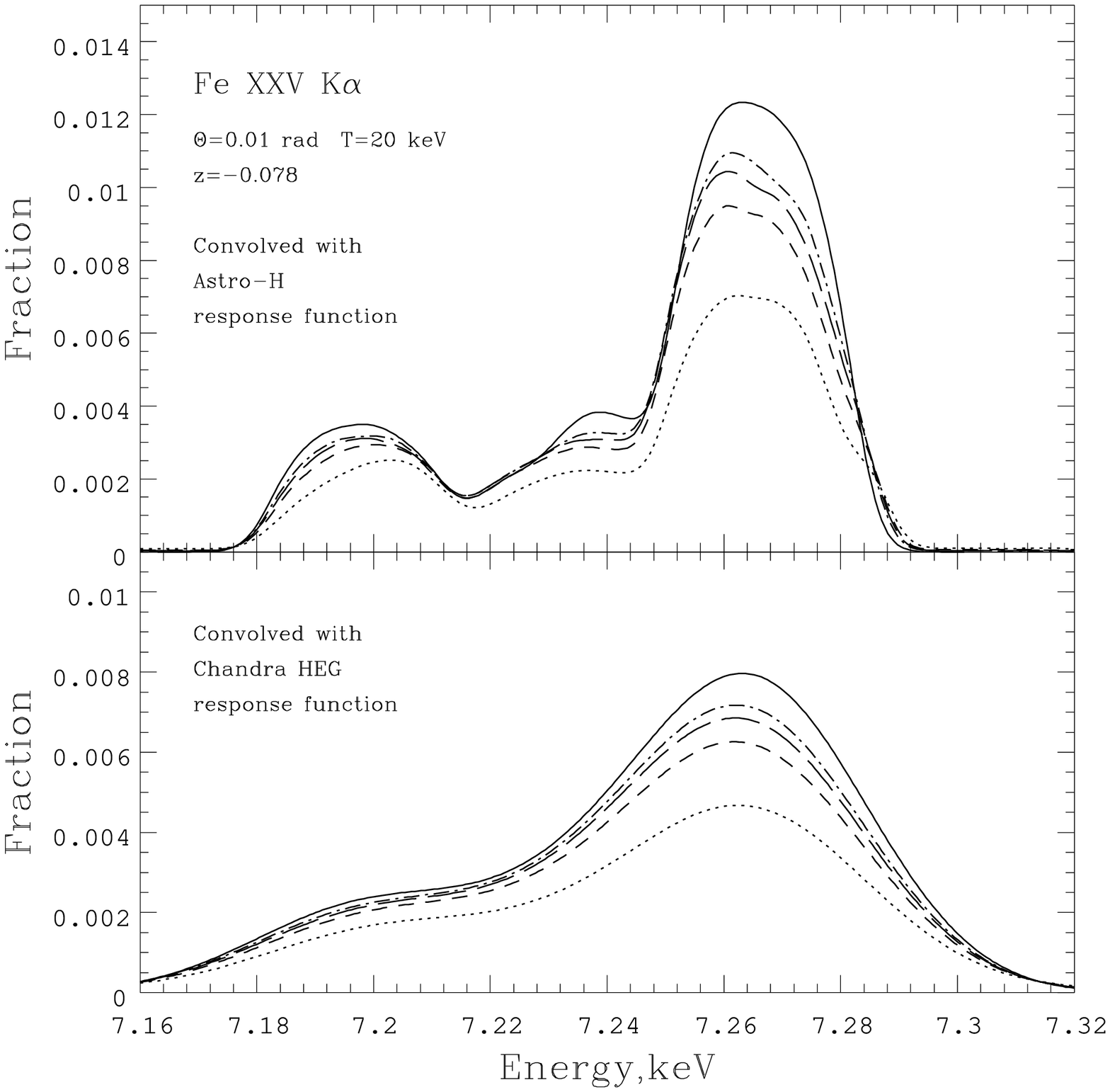}{a}{}
}
{\centering %\leavevmode
\smfigure{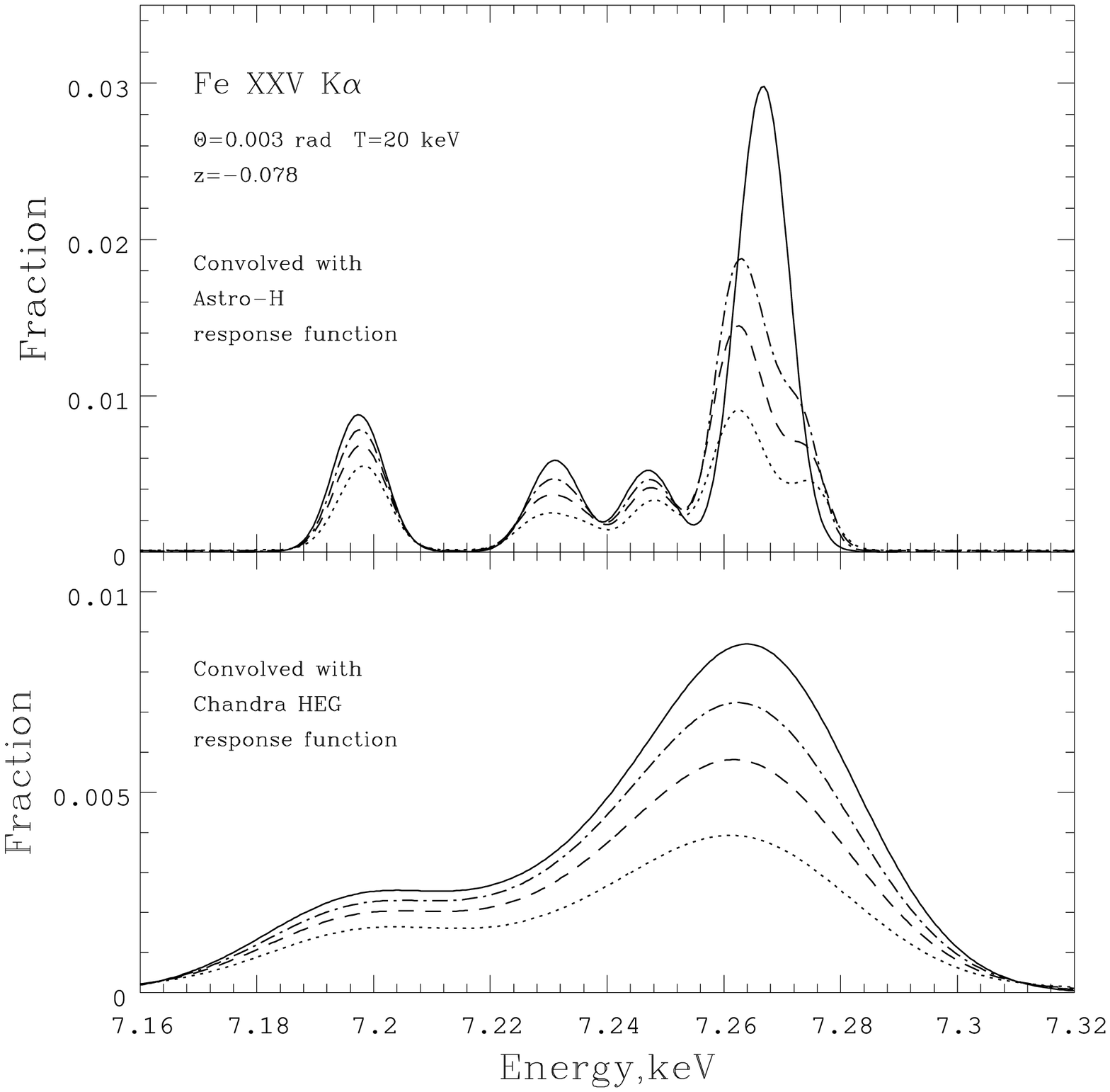}{b}{}
}

\caption{ Results of our simulations for the FeXXV $K_{\alpha}$ triplet for the approaching jet with $T_{0}=20$ keV at a phase corresponding
to $ z_{b}=-0.078 $. (a)  $ \Theta=0.01$; the upper panel shows the spectrum convolved with the Astro-H model response
function; the lower panel show the spectrum convolved with the Chandra HEG model response function; the dotted curve
for r$ r_{0}=4\times10^{10}$ cm, the curve with short dashes for $ r_{0}=8\times10^{10}$ cm, the curve with long dashes for $ r_{0}=1.8\times10^{11}$ cm, the dash-dotted curve for $ r_{0}=3.2\times10^{11}$ cm, and the solid curve indicates the unscattered line profile (the same for all $ r_{0}$).
(b) $ \Theta=0.003$ rad; the upper panel shows the spectrum convolved with the Astro-H model response function; the lower panel
shows the spectrum convolved with the Chandra HEG model response function; the dotted curve for $ n_{e0}=1.6\times10^{15}$ cm$ ^{-3} $,
the curve with short dashes for $ n_{e0}=4\times10^{14}$cm$ ^{-3} $, the dash-dotted curve for $ n_{e0}=1.0\times10^{14}$ cm$ ^{-3} $, and the solid curve
indicates the unscattered line profile (the same for all $ n_{e0}$)}

\label{specfexxv}
\end{figure*}

The results of our analysis of the simulated spectra
in the Chandra HEG resolution are presented
in Table \ref{fexxvchandra}. 
Since $FWHM_{Ch}$ at $E_{z} \simeq $ 7 keV exceeds
the expected line $FWHM_{0}=\sqrt{3}\gamma\beta\Theta \sin\chi \ E_{z}$
\citep{marshetal02} ($ \chi\approx 65^{\circ} $ is the angle between
the jet axis and the observer's direction) for $\Theta \simeq 0.01$ rad (see Fig. \ref{specfexxv}), we fitted the spectrum convolved
with the model response function by \textbf{three} Gaussians
of the same FWHM but with decoupled centroids
($E_{f},E_{i},E_{r}$ in the rest frame of the jet) and amplitudes
\textit{f, i, r} (without scattering $\left(f/r\right)_{0}=0.28  $,  
$\left(i/r\right)_{0}=0.27 $  (see Table \ref{fexxvchandra}). In what follows, the tables of
results give the line width $W$ minus the instrumental
broadening (i.e., $W=\sqrt{FWHM^{2}-FWHM_{Ch}^{2}}$ and
$W=\sqrt{FWHM^{2}-FWHM_{AH}^{2}}$ in the cases of analysis
in the Chandra HEG and Astro-H resolutions, respectively).\

The attenuation coefficient ($1-\zeta$) was calculated
for the triplet as a single line with a centroid $E_{c}$
(the first moment of the photon number distribution
in energy), i.e., its boundaries were determined with
respect to the common broad wings (see Table \ref{fexxvchandra}).\

Since the spectra convolved with the Astro-H
model response function cannot be described in such
a way, we provide only some of the derived line profiles
(Fig. \ref{specfexxv}), which illustrate the characteristic features of 
a particular scattering regime. The energy $  E$ in the
observer's frame of reference is along the horizontal
axis; the fraction of all line photons with energies
in the range from $E-\Delta E/2$ to $E+\Delta E/2$, where
$\Delta E \ll FWHM$ is the spectral bin size, is along the
vertical axis.\

As has been noted above, the triplet components
at $\Theta \geq 0.01$ rad actively interact via the optical
depths, what distorts noticeably the line shape. For
$\Theta=0.02$ rad the transverse optical depth of the jet
for Thomson scattering 
$\widehat{\tau}_{T}$ turns out to be less than a
critical value $\tau_{cr}\approx 0.19 $ (see the ``Radiative Transfer
in Lines'' Section) for the resonance line photons
at $r_{0}>4\times 10^{10} $ cm; therefore, the scatterings by
electrons cause all triplet lines to be suppressed to
approximately the same degree. The fraction of the
photons that left the line $\zeta$ reaches 30\% at
 $r_{0}=4\times 10^{10} $cm.
  Note that for $\Theta=0.02$ rad the quantities
$E_{f},E_{i},E_{r}$ and \textit{f, i, r} found in our analysis lose their
physical linkage to the forbidden, intercombination,
and resonance lines, respectively, and basically have
only a descriptive significance.\

 For  $\Theta=0.01$ rad the interaction between the
components still remains significant, but the ``extension''
of the optical depth to the low-energy region
for the resonance line photons becomes its main
manifestation. In this case, 
$ \widehat{\tau}_{T}$ turns out to be greater
than $\tau_{cr}\approx 0.15$ for the resonance photons already at
$r_{0}=8\times
10^{10}$ cm. This enhances the escape into
the wings $\zeta$ and increases the relative intensities of
the forbidden (f/r) and intercombination (i/r) lines
(see Table \ref{fexxvchandra}). Thus, the fraction of the photons
that left the line turns out to be significant, but,
nevertheless, $\zeta \leq 40\%$ for the entire triplet due to
the photons emitted inside the surface layer (see
the ``Radiative Transfer in Lines'' Section) and the
contribution from the forbidden and intercombination
components. Another effect is the increase in line
width $  W$ as a result of multiple scatterings.\

For $\Theta=0.003$ rad the interaction between the
components no longer plays a significant role, while
$\widehat{\tau}_{T}$ turns out to be greater than $\tau_{cr}\approx 0.11$ for the
resonance photons already at $n_{e0} > 1\times 10^{14}$ cm$^{-3}$.
Therefore, the resonance line is suppressed noticeably
in comparison with the forbidden and intercombination
lines. The existence of a transverse velocity
gradient has an effect on the strong asymmetry of the
resonance line profile, in agreement with the predictions
made in the ``Radiative Transfer in Lines'' Section.
The additional broadening also becomes
more pronounced.\ 
  
  In the cylindrical case of $\Theta=0.0001$ rad, the
transverse velocity gradient is negligible and almost
all of the resonance line photons produced deep in the
jet are ultimately scattered by an electron and fall into
the broad line wings. Thus, in a sense, a pure case
is realized, i.e., only the photons emitted in the jet
surface layer remain at the resonance line center. As
a result, the entire triplet noticeably loses in intensity,
while the relative contribution of the forbidden and
intercombination lines increases. In this case, the
width of the lines exceeds that of the unscattered lines
at $\Theta=0.003$ rad and is less than that of the scattered
lines at $\Theta=0.01$ rad by only a factor of 1.5. Given
the additional possibilities for broadening (see Section
5.3), this indicates that $\Theta$ is difficult to measure
accurately based on Chandra HETGS observations
of the helium-like iron $K_\alpha$ triplet. In turn, Astro-H
is ideally suited for studying the fine structure of this
line and diagnosing the jet parameters from it.\

\begin{table*}[ht]
\topcaption{Analysis of the simulation results for the FeXXV $K_\alpha$ triplet in the Chandra resolution for the approaching jet
with $ T_{0}=20$ keV at a phase corresponding to $ z_{b}=-0.078 $. The widths of the unscattered lines are $W_{0}=37.8$ eV at $ \Theta=0.02 $ rad, $W_{0}=24.0$ eV at $ \Theta=0.01 $ rad, $W_{0}=13.3$ eV at
$ \Theta=0.003 $ rad, and $W_{0}=11.9$ eV at $ \Theta=0.0001 $ rad}
\centering
\begin{tabular}{ccccccccc}\hline\hline
{}&
{1-$\zeta$}&{E$_{c}$, keV}&{W, eV}&{f/r}&{i/r}
&{E$_{f}$, keV}&{E$_{i}$, keV}&{E$_{r}$, keV}
\\[2mm]\hline
{$r_{0}, 10^{11} $ cm}&\multicolumn{8}{c}{$\Theta$=0.02 rad }\\[2mm]\hline
5.0 & 0.95 & 6.6835 & 38.2 & 0.315 & 0.630 & 6.6336 & 6.6736 & 6.7048\\
3.2 & 0.93 & 6.6835 & 38.2 & 0.310 & 0.612 & 6.6340 & 6.6727 & 6.7047\\
1.8 & 0.91 & 6.6833 & 38.3 & 0.303 & 0.600 & 6.6339 & 6.6716 & 6.7047\\
0.8 & 0.85 & 6.6833 & 38.1 & 0.287 & 0.587 & 6.6340 & 6.6695 & 6.7051\\
0.4 & 0.69 & 6.6838 & 37.0 & 0.252 & 0.563 & 6.6340 & 6.6661 & 6.7063\\
\hline
{$r_{0}, 10^{11} $ cm}&\multicolumn{8}{c}{$\Theta$=0.01 rad }\\[2mm]\hline 
5.0 & 0.95 & 6.6876 & 25.2 & 0.290 & 0.276 & 6.6371 & 6.6721 & 6.6992\\
3.2 & 0.93 & 6.6852 & 25.5 & 0.293 & 0.274 & 6.6374 & 6.6721 & 6.6990\\
1.8 & 0.88 & 6.6831 & 25.9 & 0.298 & 0.274 & 6.6378 & 6.6721 & 6.6989\\
0.8 & 0.81 & 6.6829 & 26.1 & 0.307 & 0.275 & 6.6384 & 6.6721 & 6.6989\\
0.4 & 0.62 & 6.6826 & 26.2 & 0.338 & 0.281 & 6.6397 & 6.6730 & 6.6995
\\
\hline
{$n_{e0}, 10^{14} $ cm$^{-3}$}&\multicolumn{8}{c}{$\Theta$=0.003 rad }\\[2mm]\hline 
1  &  0.86 & 6.6824 & 15.9 & 0.311 & 0.300 & 6.6372 & 6.6731 & 6.6985\\
4  &  0.71 & 6.6813 & 16.8 & 0.350 & 0.333 & 6.6376 & 6.6741 & 6.6984\\
16 & 0.51 & 6.6800 & 17.3 & 0.438 & 0.408 & 6.6381 & 6.6758 & 6.6990
\\
\hline
{$n_{e0}, 10^{14} $ cm$^{-3}$}&\multicolumn{8}{c}{$\Theta$=0.0001 rad }\\[2mm]\hline 
1   & 0.80 & 6.6820 & 14.7 & 0.338 & 0.334 & 6.6370 & 6.6733 & 6.6994\\
4   & 0.64 & 6.6803 & 15.4 & 0.406 & 0.392 & 6.6372 & 6.6745 & 6.6996\\
16 & 0.45 & 6.6781 & 15.6 & 0.515 & 0.486 & 6.6373 & 6.6758 & 6.6999
\\%[2mm]
\hline
\end{tabular}
\label{fexxvchandra}
\end{table*}

%%%% END OF  TABLES

\clearpage
%%%%%%%%%%%%%%%%%%%%%%%%%%%%%%%%%%%%%%%%%%%%
%\newpage
\subsubsection{Fe XXVI Ly$_{\alpha}$}

Since the FeXXVI Ly$_{\alpha}$
doublet corresponds to the transitions in hydrogenlike
iron, both emissivity and optical depth of the jet
plasma in these lines have a peak at slightly higher
temperatures than those in the helium-like iron K$_{\alpha}$
triplet lines (see Fig. \ref{fluxdepthprof}). This allows the Fe XXVI Ly$_{\alpha}$/ Fe XXV K$_{\alpha}$ ratio to be used to diagnose the
temperature of the hottest parts of the jet \citep{kotani96}.\

In this case, the oscillator strength $  f$ for the Ly$_{\alpha}$
transitions is considerably (approximately by a factor
of 3 for the $ 1s-2p$ ($^{1}P_{3/2}$) transition and approximately
by a factor of 6 for the $ 1s-2p$ ($^{1}P_{1/2}$) transition;
see Table \ref{linelist}) smaller than that for the resonance
component of the K$_{\alpha}$ triplet, whence, at first glance,
the doublet lines can be assumed to be less subjected
to the scattering effects inside the jet. However, the
Doppler energy shift due to the transverse jet plasma
velocity component at $\Theta\sim 0.01$ rad allows the photons
of the higher-energy doublet component to be
efficiently scattered on the transition corresponding to
the low-energy component. As a result, the scattered
line for $\Theta= 0.01$ rad has a completely different profile
compared to the initial one (see Fig. 6); for its description
in the Astro-H resolution, we used a model
consisting of two Gaussians of the same FWHM with
amplitudes A1 and A2 and centroids shifted by 
$\delta z_{1}$ and $\delta z_{2}$ relative to the initial positions for the low- and
high-energy components, respectively (Table  \ref{fexxviah}).\
  
\begin{figure*}[hb]
{\centering \leavevmode
\smfigure{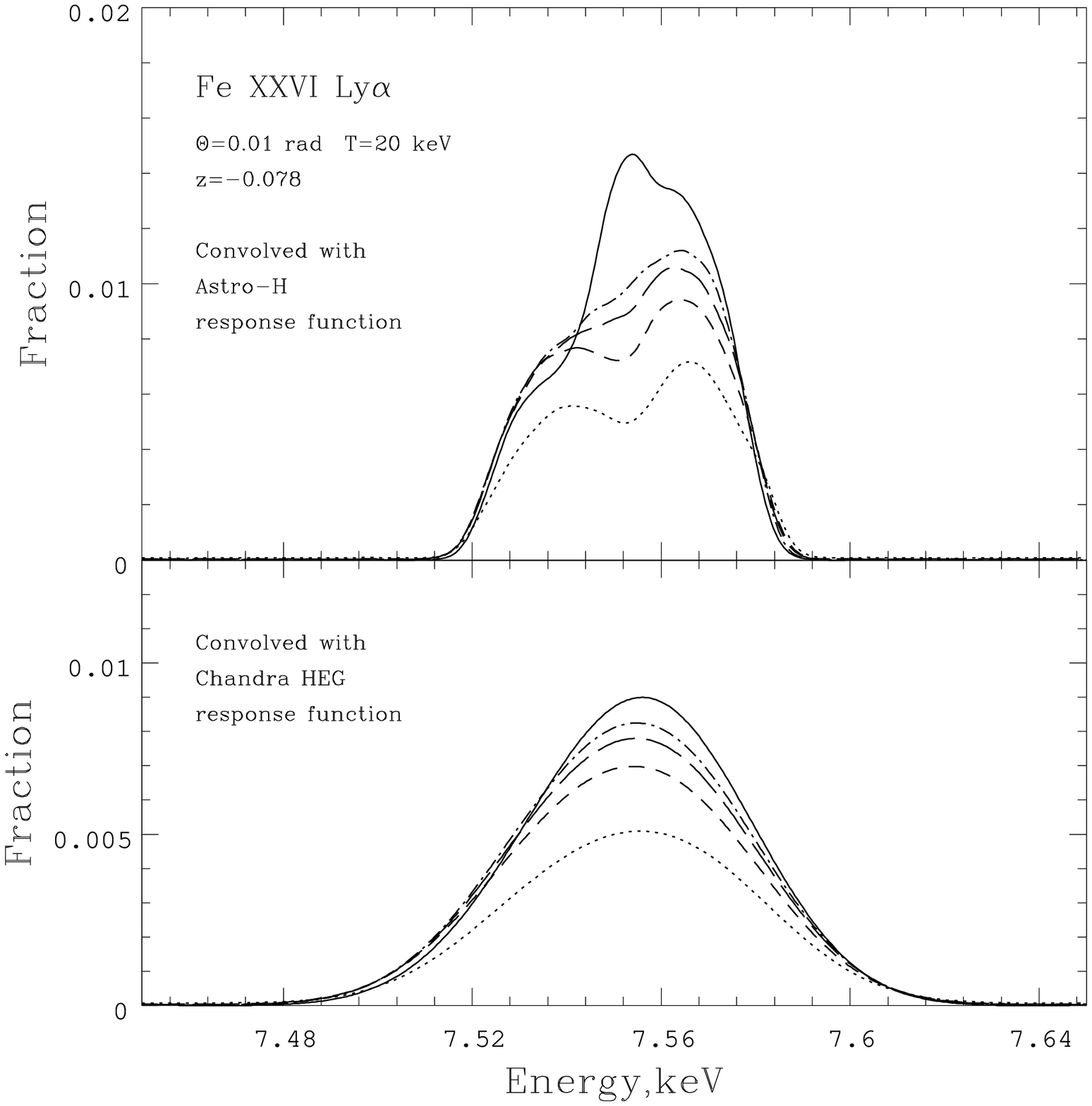}{a}{}
}
{\centering \leavevmode
\smfigure{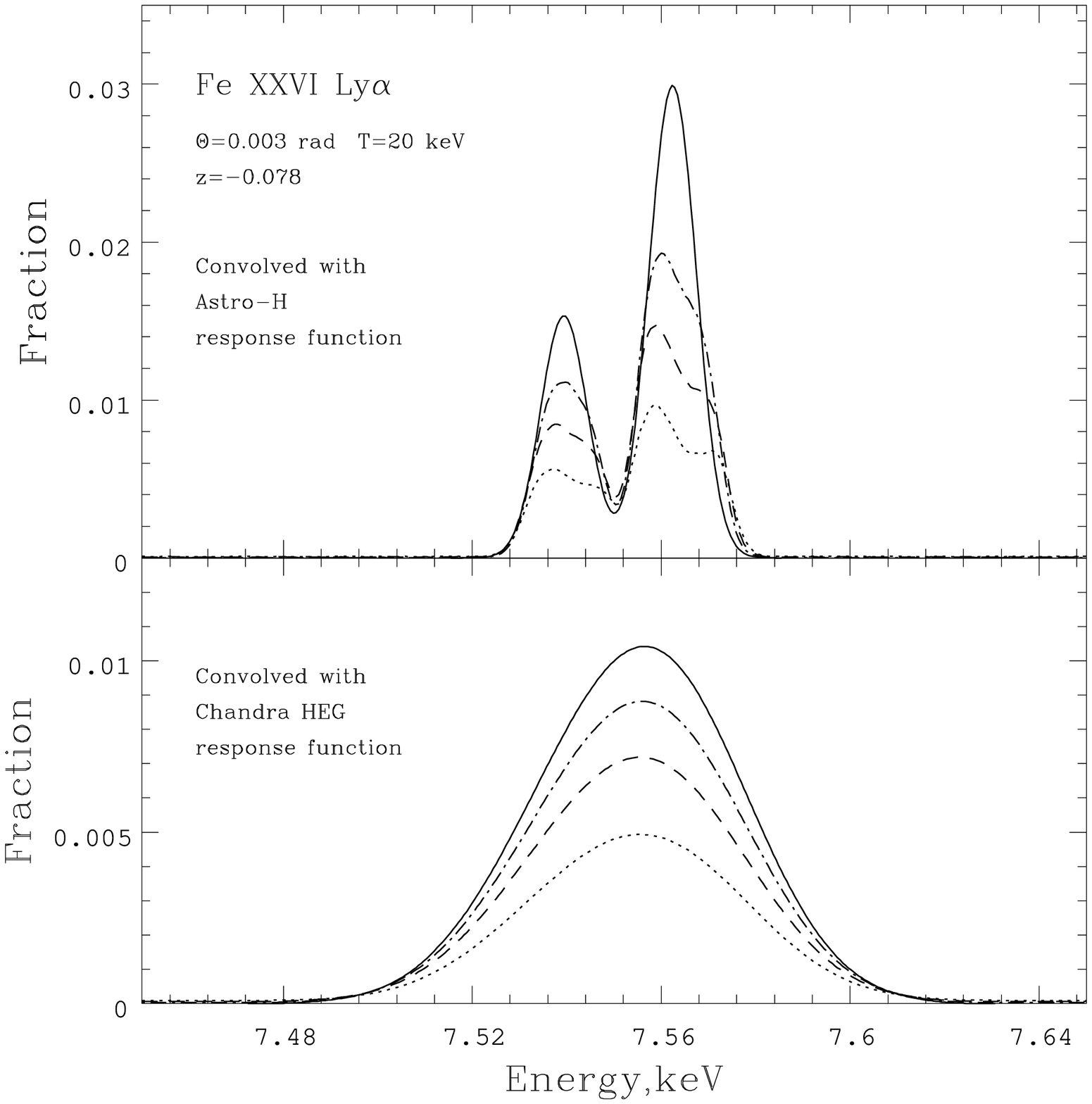}{b}{}
}

\caption{ Results of our simulations for the FeXXVI $Ly_{\alpha}$ doublet for the approaching jet with $T_{0}=20$ keV at a phase corresponding
to $ z_{b}=-0.078 $. (a)  $ \Theta=0.01$; the upper panel shows the spectrum convolved with the Astro-H model response
function; the lower panel show the spectrum convolved with the Chandra HEG model response function; the dotted curve
for r$ r_{0}=4\times10^{10}$ cm, the curve with short dashes for $ r_{0}=8\times10^{10}$ cm, the curve with long dashes for $ r_{0}=1.8\times10^{11}$ cm, the dash-dotted curve for $ r_{0}=3.2\times10^{11}$ cm, and the solid curve indicates the unscattered line profile (the same for all $ r_{0}$).
(b) $ \Theta=0.003$ rad; the upper panel shows the spectrum convolved with the Astro-H model response function; the lower panel
shows the spectrum convolved with the Chandra HEG model response function; the dotted curve for $ n_{e0}=1.6\times10^{15}$ cm$ ^{-3} $,
the curve with short dashes for $ n_{e0}=4\times10^{14}$cm$ ^{-3} $, the dash-dotted curve for $ n_{e0}=1.0\times10^{14}$ cm$ ^{-3} $, and the solid curve
indicates the unscattered line profile (the same for all $ n_{e0}$)}
\label{specfexxvi}
\end{figure*}

 When analyzing the spectra in the Chandra HEG
resolution, we used a model of a single Gaussian
with FWHM with a centroid corresponding to the
weighted mean energy of the unscattered doublet
(Table \ref{fexxvichandra}). The attenuation coefficient ($1-\zeta$) is given
for the doublet as a single line. Typical scattered line
profiles are shown in Fig. \ref{specfexxvi}. \ 

Although for $\Theta=0.02$ rad and $\Theta=0.01$  rad 
$\widehat{\tau}_{T}$
is less than $\tau_{cr}$ for the photons of the high-energy
component at all $r_{0}$, except for $r_{0}=4\times 10^{10}$ cm, the
total doublet intensity decreases considerably. This is
probably a result of the above-mentioned extension of
the optical depth to the red region for the photons of
the high-energy component. This effect is clearly illustrated
by the results of our analysis of the scattered
doublet profiles in the Astro-H resolution (Table \ref{fexxviah}).
The increase of A$ _{1} $/A$ _{2} $ with decreasing $r_{0}$ allows the
``flattening'' of the doublet profile to be judged, while
the centroid shifts $\delta z_{1}$ and $ \delta z_{2}$ point to the absence
of a physical linkage of the fitting Gaussians to the
doublet components. In addition, the effective line
broadening turns out to be very significant (Table \ref{fexxvichandra}).

For $\Theta=0.003$ rad and $\Theta=0.0001$ rad, in the
absence of any interaction between the components,
the scattering pattern is completely identical to that
described in detail for the $K_{\alpha}$ triplet. In this case, the
scattered line widthW from our analysis in the Chandra
HEG resolution turns out to be approximately
equal to the unscattered line width W$ _{0} $ for $\Theta=0.01$ rad.

Thus, owing to the interaction between the components,
the influence of scattering effects turns out
to be significant (primarily on the line profiles) even at
large $ \Theta $  almost on the entire grid of parameters. This
makes the fine structure of the $Ly_{\alpha}$ doublet profile a
very sensitive tool for diagnosing the hottest parts of
the jet, in particular, based on Astro-H observations.

%%%% BEGIN OF TABLES

\begin{table*}[ht]
\topcaption{ Analysis of the simulation results for the FeXXVI $Ly_\alpha$ doublet in the Chandra resolution for the approaching jet
with $ T_{0}=20$ keV corresponding to $ z_{b}=-0.078 $. The widths of the unscattered lines are $W_{0}=54.7$ eV at $ \Theta=0.02 $ rad , $W_{0}=35.9$ eV at $ \Theta=0.01 $ rad , $W_{0}=29.6$ eV at $ \Theta=0.003 $ rad , $W_{0}=29.5$ eV at $ \Theta=0.0001 $ rad .}
\centering
\begin{tabular}{ccccc}\hline\hline
%\\[2mm]\hline
\multirow{2}*{$r_{0}, 10^{11} $ cm}&\multicolumn{2}{c}{1-$\zeta$}
&\multicolumn{2}{c}{W, eV}\\
{}&{$\Theta$=0.02 rad }&{$\Theta$=0.01 rad }&{$\Theta$=0.02 rad }&{$\Theta$=0.01 rad }
\\[2mm]\hline
{5.0}&{0.93}&{0.93}&{59.2}&{40.3}\\
{3.2}&{0.91}&{0.91}&{61.1}&{40.9}\\
{1.8}&{0.88}&{0.86}&{62.3}&{41.8}\\
{0.8}&{0.82}&{0.78}&{64.5}&{43.7}\\
{0.4}&{0.65}&{0.58}&{68.3}&{45.3}\\
\hline
{$n_{e0}, 10^{14} $ cm$^{-3}$}&{$\Theta=3\times10^{-3}$ rad } &{$\Theta=10^{-4}$rad }&{$\Theta=3\times10^{-3}$ rad } &{$\Theta=10^{-4}$ rad }
\\[2mm]\hline
{1.0}&{0.86}&{0.82}&{30.9}&{31.1}\\
{4.0}&{0.71}&{0.66}&{31.8}&{32.0}\\
{16.0}&{0.49}&{0.45}&{32.9}&{33.0}\\
\hline
\end{tabular}
\label{fexxvichandra}
\end{table*}

\begin{table}[hb]
\topcaption{Analysis of the simulation results for the FeXXVI $Ly_\alpha$ doublet in the Astro-H resolution for the approaching jet
with $ T_{0}=20$ keV corresponding to $ z_{b}=-0.078 $.}
\centering
\begin{tabular}{ccccc}
\multicolumn{5}{c}{$\Theta=0.01$ rad }\\
\hline\hline
{$r_{0}, 10^{11} $ cm}&{A1/A2}&{W, eV}&{$\delta z_{1}$,  
$10^{-4}$}&{$\delta z_{2}, 10^{-4}$}\\[2mm] 
\hline
5.0 & 0.66 & 26.18 & 1.46 & -1.69\\
3.2 & 0.69 & 25.73 & 1.34 & -2.17\\
1.8 & 0.73 & 25.57 & 1.59 & -2.66\\
0.8 & 0.78 & 24.96 & 2.20 & -3.52\\
0.4 & 0.80 & 24.78 & 1.48 &-5.20\\ \hline
\end{tabular}
\label{fexxviah}
\end{table}
%%%% END OF FeXXVI TABLES

%\clearpage
%%%%%%%%%%%%%%%%%%%%%%%%%%%%%%%%%%%%%%%%%%%%
%\newpage
\subsubsection{Ni XXVII K$_{\alpha}$ +Fe XXV K$_{\beta}$}
%%%%%%%%%%%%%%%%%%%%%%%%%%%%%%%%%%%%%%%%%%%

\begin{figure}[htb]
\centering
\includegraphics[width=1.0\columnwidth]{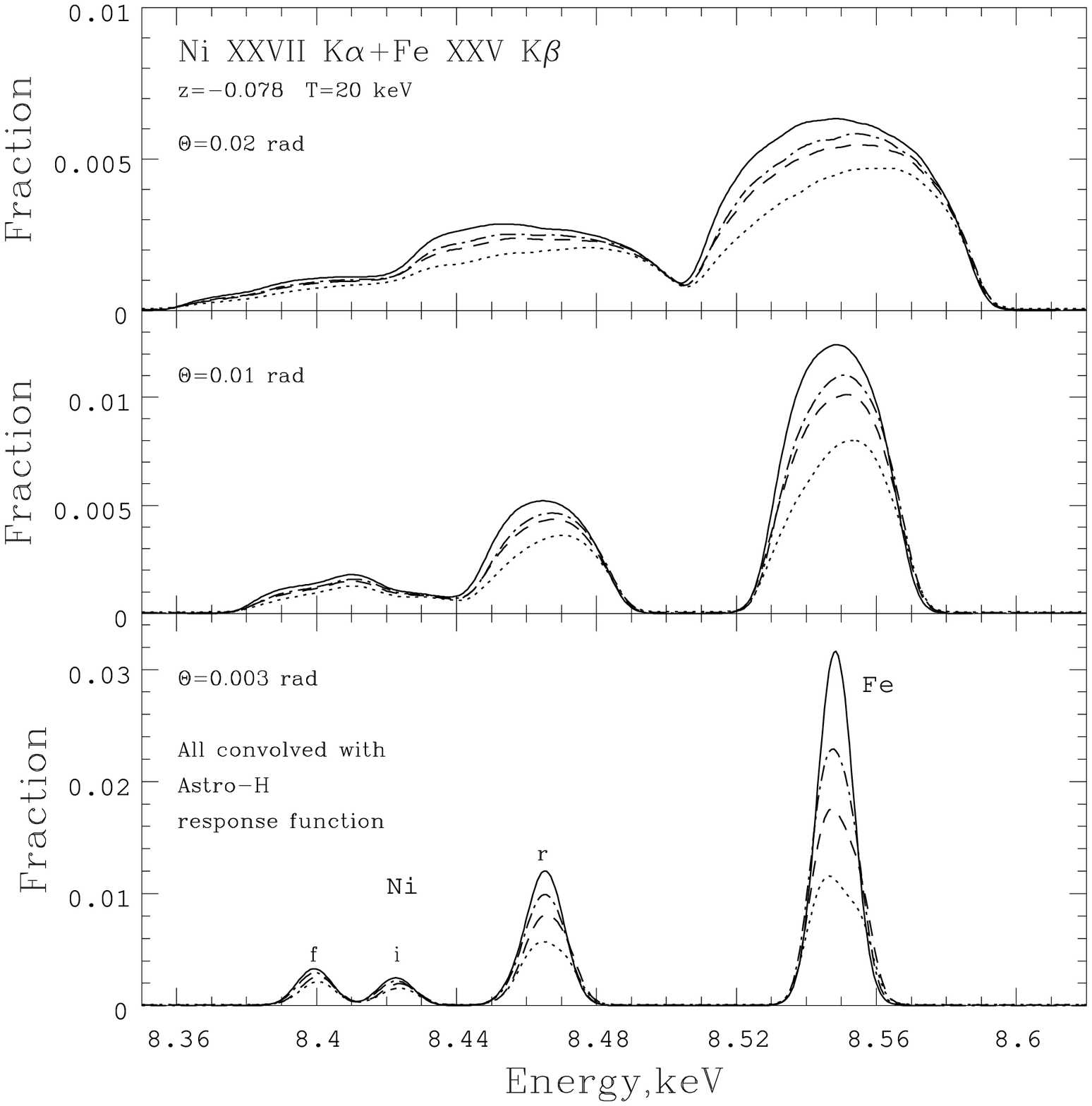}
\caption{Results of our simulations for the Ni XXVII $K_{\alpha}$ and Fe XXV $K_{\beta}$ for the approaching jet with $T_{0}=20$ at a
phase corresponding to $ z_{b}=-0.078 $.
(a) The spectrum convolved with the Astro-H model response function for $ \Theta=0.02$ rad;
(b) the spectrum convolved with the Astro-H model response function for $ \Theta=0.01$ rad; the dotted curve for $ r_{0}=4\times10^{10}$ cm,
the curve with short dashes for $ r_{0}=8\times10^{10}$ cm, the dash-dotted curve for $ r_{0}=1.8\times10^{11}$ cm, and the solid curve indicates
the unscattered line profile (the same for all $ r_{0}$); (c) the spectrum convolved with the Astro-H model response function for
$ \Theta=0.003$ rad; the dotted curve for $ n_{e0}=1.6\times10^{15}$ cm $ ^{-3} $, the curve with short dashes for $ n_{e0}=4\times10^{14}$ cm $ ^{-3} $, the dash-
doted curve for $ n_{e0}=1.0\times10^{14}$ cm $ ^{-3} $, and the solid curve indicates the unscattered line profile (the same for all $ n_{e0}$).}
\label{specnife}
\end{figure}

The currently
available observations near the nickel $K_{\alpha}$ triplet point
to an excess of radiation compared to that expected
from the FeXXV (6.7 keV) line intensity within the
standard model with solar elemental abundances (Table.
\ref{GaugeTableX}). This is interpreted in terms of a nickel overabundance
in the jet plasma (\cite{brinketal05},
\cite{medved10}). 
Since present-day instruments
do not allow NiXXVII $K_{\alpha}$ and FeXXV $K_{\beta}$
to be reliably resolved, we performed joint simulations
of these lines, although a noticeable overlap takes
place only for $\Theta=0.02$ rad (see Fig. \ref{specnife}).\

We do not provide the results of our analysis of
the simulated spectra in the Chandra HEG resolution
due to the low sensitivity of this instrument near $ E_{z}\simeq 8.5 $ keV.\

When analyzing the spectra in the Astro-H resolution,
we used a model consisting of five Gaussians
of the same FWHM with decoupled amplitudes but
fixed centroids. The attenuation coefficients ($1-\zeta$) are
determined separately for the nickel triplet and K$\beta$ of
helium-like iron (see Table \ref{nifeah}).\

Whereas 
$\widehat{\tau}_{T}$ is much less than $\tau_{cr}$ for the photons of
the nickel resonance line on the entire grid of quasiadiabatic
model parameters, $\tau_{cr}$ is not so large for the
photons of the iron $K_\beta$ line and even turns out to be
less than 
$\widehat{\tau}_{T}$ at $\Theta=0.01$ rad and $r_{0}=4\times 
10^{10}$ cm. In
particular, this is reflected in different dependences of
$\zeta$ on $r_{0}$ (see Table ref{nifeah}). As a result, the F(Ni)/F(Fe)
ratio slightly changes, but this change takes place
only at $r_{0}=4\times 10^{10}$ cm and does not exceed 10\%
in magnitude. Thus, the main effect is a general decrease
in intensity due to Thomson scattering almost
on the entire grid of quasi-adiabatic model parameters.\ 

\ For the quasi-cylindrical models, the overall pattern
of radiative transfer is very similar to the case of
the FeXXV $ K_{\alpha} $ triplet, given that for the resonance
line of the nickel triplet and $K_{\beta} $ of iron 
$\widehat{\tau}_{T}>\tau_{cr}$ only
for $\alpha>30$ and $\alpha>10$, respectively (see Table \ref{GaugeTableX}).\

Thus, the intensity ``transfer'' between the iron
$K_\beta$ line and the nickel $K_{\alpha}$ triplet through resonant
scattering turns out to be fairly likely. At the same
time, the intensity of the entire set of lines does
not undergo such a dramatic decrease as the iron
$K_{\alpha} $ triplet. However, this cannot completely explain
the observed excess of radiation in this region relative
to that expected from the FeXXV $ K_\alpha $ (6.7 keV) line
under the assumption of solar elemental abundances.\
   
  Astro-H observations allow the nickel triplet lines
and $K_{\beta}$ of iron to be reliably resolved, thereby making
it possible to determine the nickel abundance relative
to iron in the jets of SS 433 \textit{in situ}.

\begin{table}[hb]
\topcaption{ Analysis of the simulation results for the Ni XXVII K$_{\alpha}$ +Fe XXV K$_{\beta}$ in the Astro-H resolution for the approaching jet
with $ T_{0}=20$ keV corresponding to $ z_{b}=-0.078 $.
 The widths of the unscattered lines are $W_{0}=30.3$ eV at $ \Theta=0.01 $ rad , $W_{0}=10.8$ eV at $ \Theta=0.003 $ rad , $W_{0}=7.65$ eV at $ \Theta=0.0001 $ rad .}
\centering
\begin{tabular}{ccccc}\hline\hline
%\\[2mm]\hline
\multirow{2}*{}&
{1-$\zeta$(Ni)}&{1-$\zeta$(Fe)}&{F(Ni)/F(Fe)}&{W, eV}
\\[2mm]\hline
{$r_{0}, 10^{11} $ cm }&\multicolumn{4}{c}{$\Theta$=0.01 rad }\\[2mm]\hline 
5.0 & 0.94 & 0.94 & 0.61 & 30.3\\
3.2 & 0.92 & 0.92 & 0.61 & 30.6\\
1.8 & 0.89 & 0.89 & 0.61 & 30.9\\
0.8 & 0.84 & 0.83 & 0.61 & 31.3\\
0.4 & 0.71 & 0.65 & 0.65 & 31.3
\\%[2mm]
\hline
{$n_{e0}, 10^{14} $ cm $^{-3}$}&\multicolumn{4}{c}{$\Theta$=0.003 rad }\\[2mm]\hline 
1  & 0.90 & 0.88 & 0.62 & 13.2\\
4  & 0.80 & 0.74 & 0.65 & 14.7\\
16 & 0.63 & 0.53 & 0.72 & 16.2
 
\\%[2mm]
\hline
{$n_{e0}, 10^{14} $ cm $^{-3}$}&\multicolumn{4}{c}{$\Theta$=0.0001 rad }\\[2mm]\hline 
1  & 0.87 & 0.83 & 0.64 & 11.2\\
4  & 0.77 & 0.68 & 0.70 & 12.6\\
16 & 0.60 & 0.48 & 0.77 & 14.1
\\%[2mm]
\hline
\end{tabular}
\label{nifeah}
\end{table}

%%%%%%%%%%%%%%%%%%%%%%%%%%%%%%%%%%%%%%%%%%%
%\newpage
\subsubsection{S XV K$_{\alpha}$}	
\label{sxvrg}
\begin{figure*}[hb]

{\centering \leavevmode
\smfigure{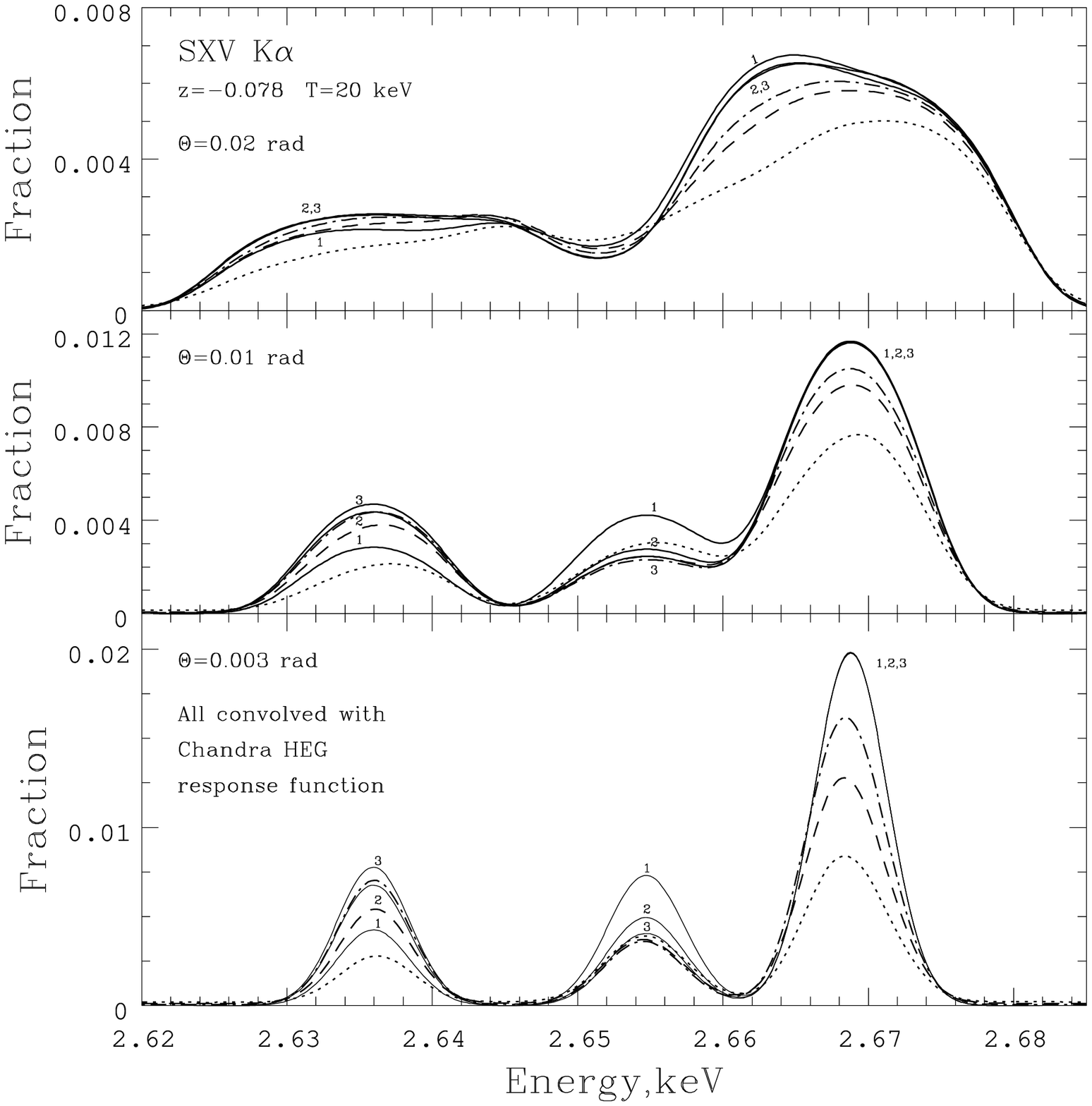}{a}{}
}
{\centering \leavevmode
\smfigure{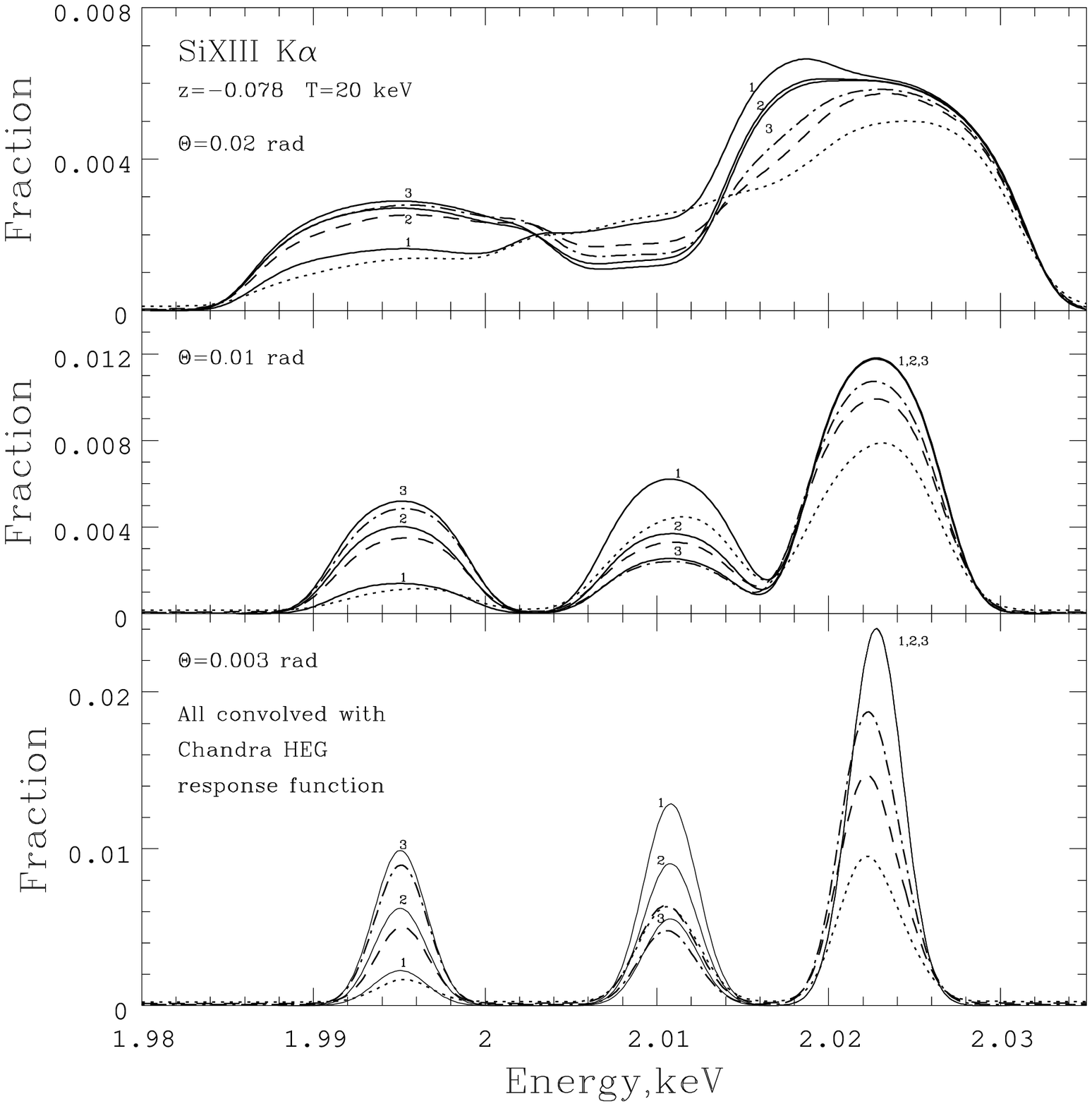}{b}{}
}

\caption{Results of our simulations for the SXV $K_{\alpha}$ triplet (a) and the SiXIII $K_{\alpha}$ triplet (b) for the approaching jet with $T_{0}=20$ keV at a phase corresponding to $ z_{b}=-0.078 $. The upper panel: $ \Theta=0.02$ rad; the spectrum convolved with the Chandra
HEG model response function; the dotted curve for $ r_{0}=4\times10^{10}$ cm, the curve with short dashes for $ r_{0}=8\times10^{10}$cm, the
dash-dotted curve for $ r_{0}=1.8\times10^{11}$  cm, 1, 2, 3 represent the unscattered line profile for $ r_{0}=4\times10^{10}$ cm, $ r_{0}=8\times10^{10}$ cm,
and $ r_{0}=1.8\times10^{11}$ cm, respectively. The middle panel: $ \Theta=0.01$ rad; the spectrum convolved with the Chandra HEG
model response function; the dotted curve for $ r_{0}=4\times10^{10}$ cm, the curve with short dashes for $ r_{0}=8\times10^{10}$  cm, the dash-dotted curve for $ r_{0}=1.8\times10^{11}$ cm, 1, 2, 3 represent the unscattered line profile for $ r_{0}=4\times10^{10}$ cm, $ r_{0}=8\times10^{10}$  cm,
and $ r_{0}=1.8\times10^{11}$ cm, respectively. The lower panel: $ \Theta=0.003$ rad; the spectrum convolved with the Chandra HEG
model response function; the dotted curve for $ n_{e0}=1\times10^{14}$ cm $ ^{-3} $, the curve with short dashes for $ n_{e0}=4\times10^{14}$ cm $ ^{-3} $,
the dash-dotted curve for $ n_{e0}=1.6\times10^{15}$ cm $ ^{-3} $, 1, 2, 3 represent the unscattered line profile for $ n_{e0}=1.6\times10^{15}$ cm $ ^{-3} $,
$ n_{e0}=4\times10^{14}$ cm $ ^{-3} $, and $ n_{e0}=1\times10^{14}$ cm $ ^{-3} $, respectively.}
\label{specsxvsixiii}
\end{figure*}

The overall pattern typical of
the K$_{\alpha}$ triplets of helium-like ions at low energies
(we will be interested primarily in sulfur and silicon;
see Table  \ref{GaugeTableX}) differs from that for high-energy triplets
(e.g., of iron) in several respects. First, the triplet
components have a large relative spectral separation
($\Delta E /\Delta E_{D}$); as a result, the overlap and, hence, the
interaction take place only for large $\Theta \sim 0.02$ rad (see
Fig. \ref{specsxvsixiii}a). Nevertheless, we provide the attenuation
coefficients $1-\zeta$ for the entire triplet (see Table \ref{sxvchandra}).
Second, $n_{e,crit}\propto Z^{13} $ turns out to be $10^{14}$ cm $^{-3}$
(an accurate calculation for the corresponding temperature
range was performed using the Cloudy 08.00 
code). Therefore, for the fine structure of the triplets
to be simulated in detail, the intensity redistribution of
the forbidden and intercombination lines as a result of
the collisional excitations from $^{3}S_{1}$ (the upper level of
the forbidden component) to $^{3}P_{0,1,2}$ (the upper levels
of the intercombination components) should be taken
into account \citep{PD10}. This redistribution
is described in terms of their intensity ratio 
$R\left(n_{e}\right)=z/(x+y)=\frac{r_{0}}{1+n_{e}/n_{e,crit}\left(T\right)}$,
 where $r_{0}$ is the value
of this ratio in the low-density limit (from APEC), and
the dependence $n_{e,crit}\left(T\right)$ is obtained by fitting the
Cloudy computation results.\

Since Astro-H is inferior to Chandra HETGS in
its spectral characteristics near the brightest lines of
hydrogen- and helium-like sulfur and silicon, we omit
the results of our analysis of the simulated spectra
in the Astro-H resolution. For our analysis in the
Chandra HEG resolution (Fig. \ref{specsxvsixiii}a) we used, where
possible, a model consisting of three Gaussians of the
same FWHMwith fixed centroids but with decoupled
amplitudes w, x + y, and z for the resonance, intercombination,
and forbidden components, respectively.
Since the triplet components have different
characteristics with respect to the scattering effects,
one might expect these effects to affect noticeably
the ratio R = z/(x + y) and the ratio G(T) = (z +
x + y)/w of the total intensity of the forbidden and
intercombination lines to the intensity of the resonance
line, which is used to diagnose the plasma
temperature (see, e.g., \cite{PD10}). Given
the importance of these ratios from the viewpoint of
the physical interpretation of observational data, the
summary table of results provides precisely the values
of R and G (see Table  \ref{sxvchandra}).\

\begin{table*}[hb]
\topcaption{Analysis of the simulation results for the SXV $K_\alpha$ triplet in the Chandra resolution for the approaching jet with
$ T_{0}=20$ keV at a phase corresponding to $ z_{b}=-0.078 $}
\centering
\begin{tabular}{ccccc}\hline\hline
{}&{1-$\zeta$}&{R (R$_{0}$)}&{G (G$_{0}$)}
&{W (W$_{0}$) , eV}\\[2mm] 
\hline
{$r_{0}, 10^{11} $ cm }&\multicolumn{4}{c}{$\Theta=0.01$ rad }\\[2mm]\hline
5.0 & 0.97 & 1.98 ( 1.92 )& 0.65 ( 0.64 ) & 8.92 ( 8.77 )\\
3.2 & 0.96 & 1.98 ( 1.92 )& 0.65 ( 0.64 ) & 8.96 ( 8.77 )\\
1.8 & 0.93 & 1.91 ( 1.92 )& 0.65 ( 0.64 ) & 9.06 ( 8.77 )\\
0.8 & 0.88 & 1.58 ( 1.62 )& 0.65 ( 0.64 ) & 9.09 ( 8.77 )\\
0.4 & 0.71 & 0.71 ( 0.73 )& 0.67 ( 0.64 ) & 8.97 ( 8.77 )\\
\hline
{$n_{e0}, 10^{14} $ cm $^{-3}$}&\multicolumn{4}{c}{$\Theta=0.003$ rad }\\[2mm]\hline
1  & 0.90 & 1.87 ( 1.87 ) & 0.64 ( 0.59 ) & 3.57  (2.85)\\
4  & 0.75 & 1.39 ( 1.34 ) & 0.69 ( 0.59 ) & 3.83  (2.85)\\
16 & 0.55 & 0.66 ( 0.57 ) & 0.76 ( 0.59 ) & 4.16 (2.85)\\
\hline
{$n_{e0}, 10^{14} $ cm $^{-3}$}&\multicolumn{4}{c}{$\Theta=0.0001$ rad }\\[2mm]\hline
1   & 0.80 & 1.75  ( 1.71 ) & 0.71 ( 0.59 ) & 2.93 ( 1.73 )\\
4   & 0.65 & 1.26  ( 1.14 )& 0.81 ( 0.59 ) & 3.25 ( 1.73 )\\
16 & 0.46 & 0.62  ( 0.49 )& 0.90  ( 0.59 ) & 3.52 ( 1.73 )\\
\hline
\end{tabular}
\label{sxvchandra}
\end{table*}

For $\Theta=0.02$ rad the interaction between the
triplet components resembles the situation with the
$K_\alpha$ triplet of helium-like iron at $\Theta=0.01$ rad, but
complicated by the intensity redistribution of the forbidden
and intercombination components. Therefore,
we do not present the results of a detailed analysis
for $\Theta=0.02$ rad, bearing in mind the impossibility of
their clear and unambiguous interpretation.\

For $\Theta=0.01$ rad, as a result of the rapid decrease
in $n_{e}\varpropto \xi^{-2}$ with increasing $\xi$, the effective optical
depth for scattering by free electrons 
$\widehat{\tau}_{T}$ turns out to
be greater than $\tau_{cr}$ for the resonance photons only
at $r_{0}=4\times 10^{10}$ cm, despite the large optical depth
of the jet for resonant scattering in the allowed S XV
triplet component (see Fig. \ref{fluxdepthprof}). Therefore, the overall
pattern is similar to that considered for the Ni XXVII
$K_\alpha$ triplet, i.e., the influence of scattering effects is
small, the ratios R and G differ from their unscattered
values R$_{0}  $ and G$_{0}  $ only slightly (see Table \ref{sxvchandra}).\

In the case of quasi-cylindrical models, the overall
pattern of the spectra corresponds to that for the
Fe XXV $K_\alpha$ triplet (here, $n_{e}$ changes little). As a
result, the ratios R and G as well as the effective line
width change noticeably. However, even in the case
of an increase by a factor of 1.5-2, the line width
for $ \Theta \leqslant 0.003 $ rad remains considerably smaller than
 that at $ \Theta=0.01 $ (as distinct from the analogous
situation for the FeXXV $K_\alpha$ triplet).\

%%%%%%%%%%%%%%%%%%%%%%%%%%%%%%%%%%%%%%%%%%%
%\newpage
\subsubsection{Si XIII K$_{\alpha}$}

The general scheme for analyzing
and presenting the results for the K$_{\alpha}$ triplet
of helium-like silicon is completely identical to the
scheme described in detail for the sulfur triplet in the
preceding section. Therefore, here we omit the determination
of the quantities given in Table \ref{sixiiichandra}. Similarly,
the interpretation of the results of our analysis closely
coincides with that for the sulfur triplet.\

 Nevertheless, an important distinctive feature of
the silicon triplet is the possible influence of satellites
on measured characteristics (e.g., R and G). We
will defer a detailed discussion of this question until
Section \ref{satellites}, while for now note that the diagnostics
of plasma parameters based on the ratios R and G
can be strongly affected by the scattering effects. For
example, an increase in these ratios (see Tables
 \ref{sxvchandra} and \ref{sixiiichandra}) 
 as a result of scattering relative to the values
corresponding to the optically thin approximation
\citep{PD10} must lead to an underestimation
of the density and temperature from observations. \ 
 
Given the excellent spectral resolution of Chandra
HETGS in this energy range, a detailed analysis of
the line profiles corrected for the effects considered
can provide information about the low-temperature
part of the jet.

\begin{table*}[hb]
\topcaption{Analysis of the simulation results for the SiXIII $K_\alpha$ triplet in the Chandra resolution for the approaching jet with
$ T_{0}=20$ keV at a phase corresponding to $ z_{b}=-0.078 $}
\centering
\begin{tabular}{ccccc}\hline\hline
{}&{1-$\zeta$}&{R (R$_{0}$)}&{G (G$_{0}$)}
&{W (W$_{0}$) , eV}\\[2mm] 
\hline
{$r_{0}, 10^{11} $ cm }&\multicolumn{4}{c}{$\Theta=0.01$ rad }\\\hline
5.0 & 0.98 & 2.33  ( 2.33 ) & 0.720 ( 0.713 )& 6.90 ( 6.76 )\\
3.2 & 0.96 & 2.25 ( 2.27 ) & 0.716 ( 0.709 )& 6.92 ( 6.75 )\\
1.8 & 0.94 & 1.95  ( 1.98 ) & 0.709 ( 0.699 )& 6.91 ( 6.71 )\\
0.8 & 0.89 & 1.09  ( 1.10 ) & 0.706 ( 0.691 )& 6.93 ( 6.67 ) \\
0.4 & 0.72 & 0.26 ( 0.26 )  & 0.708 ( 0.671 ) & 6.68 ( 6.56 )\\
\hline
{$n_{e0}, 10^{14} $ cm $^{-3}$}&\multicolumn{4}{c}{$\Theta=0.003$ rad }\\[2mm]\hline
1   & 0.90 & 1.79 (1.76) & 0.70 (0.64) & 2.63 (2.08)\\ %& 2.08\\\hline
4   & 0.76 & 0.73 (0.67) & 0.74 (0.64) & 2.80 (2.08)\\ %& 2.08\\\hline
16 & 0.55 & 0.22 (0.17) & 0.79 (0.64) & 2.96 (2.08)%& 2.08\\\hline
\\[1mm]\hline
{$n_{e0}, 10^{14} $ cm $^{-3}$}&\multicolumn{4}{c}{$\Theta=0.0001$ rad }\\[2mm]\hline
1   & 0.78 & 1.38 (1.24) & 0.81 (0.63) & 2.26 (1.15)\\  %& 1.15\\\hline
4   & 0.63 & 0.61 (0.48) & 0.89  (0.63) & 2.51 (1.15)\\    %& 1.15\\\hline
16 & 0.45 & 0.21 (0.14) & 0.91  (0.63) & 2.68 (1.15)\\%& 1.15\\\hline
\hline
\end{tabular}
\label{sixiiichandra}
\end{table*}

%\clearpage
%%%%%%%%%%%%%%%%%%%%%%%%%%%%%%%%%%%%%%%%%%%
%\newpage
\subsection{Broad Line Wings}
\begin{figure}[h!]
{\centering \leavevmode
\epsfxsize=1.\columnwidth \epsfbox{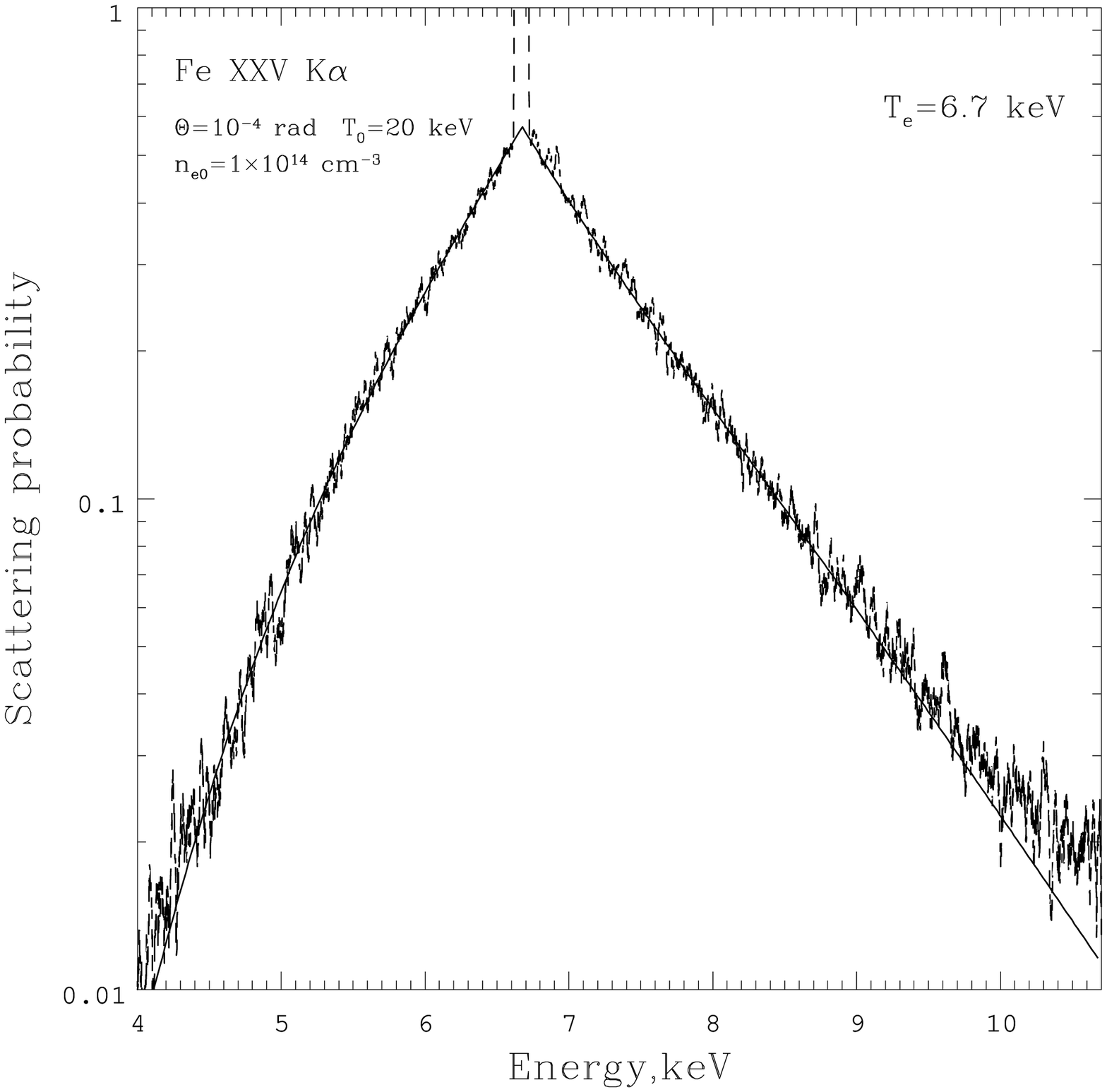}\hfill\hfil
\epsfxsize=1.\columnwidth \epsfbox{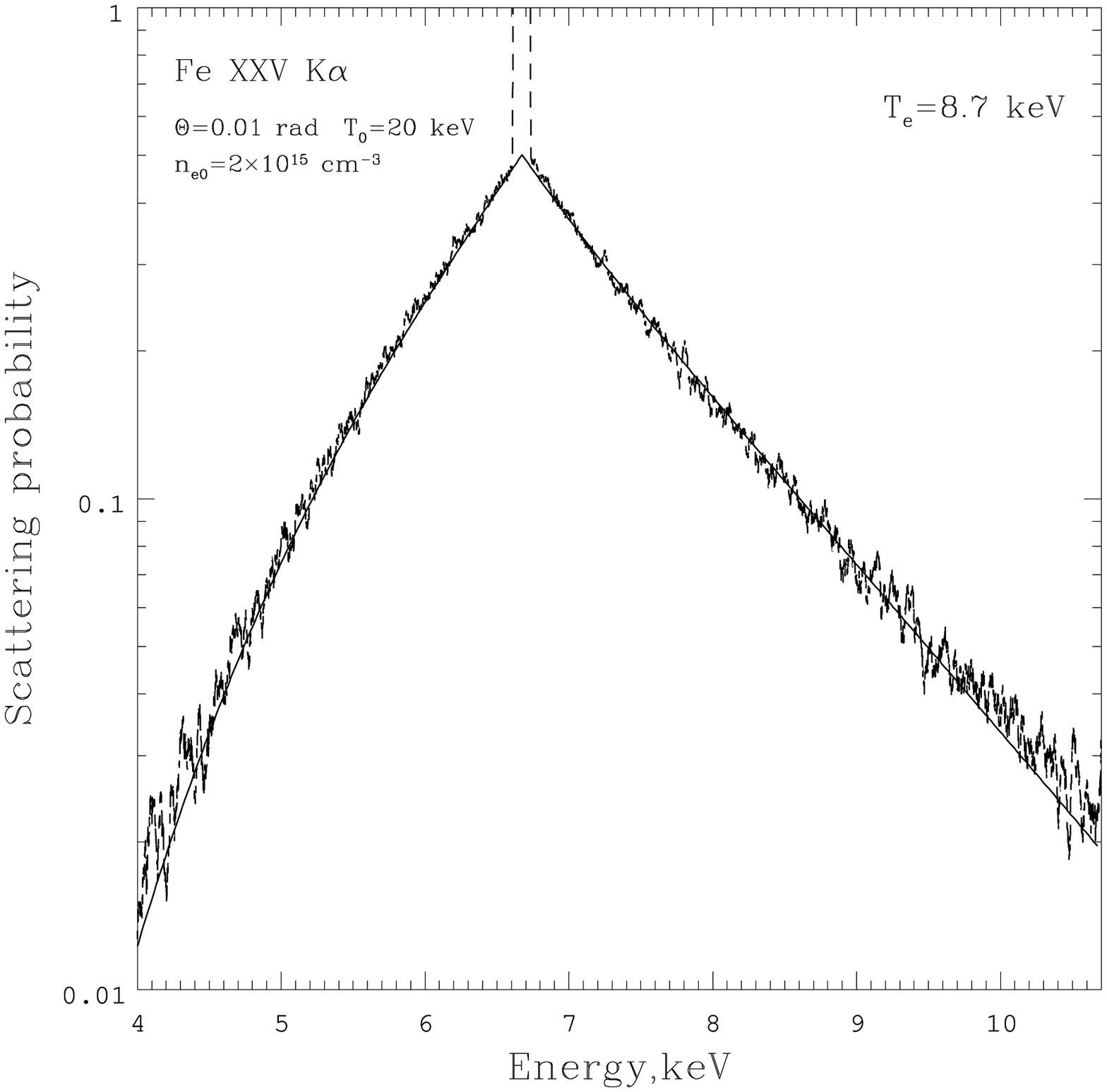}}
%\vspace{0.5cm}
\caption{Fitting the broad wings of the scattered FeXXV $K_\alpha $ line in the jet's frame of reference by the single Compton electron
scattering kernel from \cite{saz00} for one of the quasi-cylindrical models (top) and one of the quasi-adiabatic
models (bottom). The photon scattering probability for a line with the central energy $ E_{0}=6.7 $ keV and a cloud of electrons with
temperature $T_{e}$ is along the vertical axis.}
\label{kernelfits}
\end{figure}

\
Broad line wings are formed at a large number of
photons that left the line due to their scattering by a
free electron. The shape of these wings is well described
by the single Compton scattering kernel with
the application of relativistic corrections in the case
of an isotropic radiation field defined by Eq. (19) from
\cite{saz00}. In this case, the only
normalization is the quantity $\zeta$, i.e., the fraction of the
photons that left the line, while the only parameter in
fitting is the effective temperature of the scattering
electrons $T_{e}$. This temperature reflects the contribution
from various parts of the jet to the scattered
radiation and, hence, allows the temperature and density
profiles along the jet to be judged. For example,
on average, higher-temperature fits correspond to the
class of quasi-adiabatic models than to the class of
quasi-cylindrical ones (see Fig. \ref{kernelfits}). This is due to a
rapid decrease in density against the background of
a slow change in temperature for the former and the
directly opposite situation for the latter.\

This single Compton scattering kernel has the
following characteristic features: first, the presence of
a cusp near $E=E_{0}$ and, second, a noticeable asymmetry--
the ``right'' (high-energy) wing is much
broader than the ``left'' (low-energy) one. In addition,
there is a probability of photon rescattering by
an electron. However, at 
$\widehat{\tau}_{T}\sim 0.1$ the influence of rescatterings is negligible, with the possible exception
of the far edges of the wings (see Fig. \ref{kernelfits}). \

Since, in general, the broad wings of scattered
lines are \textbf{always} present in the spectrum(the question
is only in what quantities), an accurate measurement
of their shape can be used as a universal (though
model-dependent) tool for determining the jet parameters
that is already accessible for present-day instruments.
Undoubtedly, the contribution from the wings
of neighboring lines slightly complicates the problem,
but, at the same time, independent measurements
in the high- and low-energy parts of the spectrum
can reveal particular local features in the density and
temperature distributions. \
 
At the same time, it should be remembered that
the continuum itself is also subject to scattering inside
the jet, which is especially important for the
photons in resonance with the electron transitions
in ions. Nevertheless, our simulations with the addition
of an appropriate wide continuum component
showed that the continuum ``subsidence'' under the
resonance lines for lines with significant equivalent
widths affects only slightly the observed parameters
of these lines ($ \lesssim 5 \% $ for the intensity). Since a detailed
discussion of the scattering effects on the continuum
is beyond the scope of this paper, we will restrict ourselves
here only to the remark made and will consider
a number of effects affecting directly the lines.

\clearpage
%%%%%%%%%%%%%%%%%%%%%%%%%%%%%%%%%%%%%%%%%%%
\subsection{The Influence of Other Effects}
\
Apart from the scattering inside the jet, there are
also several effects capable of affecting the observational
line characteristics (intensity, width, etc.) and
the line ratios used to diagnose the plasma parameters.
In this Section, we estimate the possible effects
of satellites of bright lines, nutations, and photoionization
and photoexcitation.

\subsubsection{Satellites and the Overlap of Close
Lines}
\label{satellites}
\
On the one hand, the presence of satellites
near bright lines slightly complicates the analysis
and interpretation of the spectra, but, on the other
hand, investigating the satellite parameters (provided
an appropriate spectral resolution of the instrument)
is a powerful tool for diagnosing the parameters
of the emitting medium \textit{per se} (see, e.g., \cite{PD10}).
\

\begin{figure}[ht]
{\centering \leavevmode
\smfigure{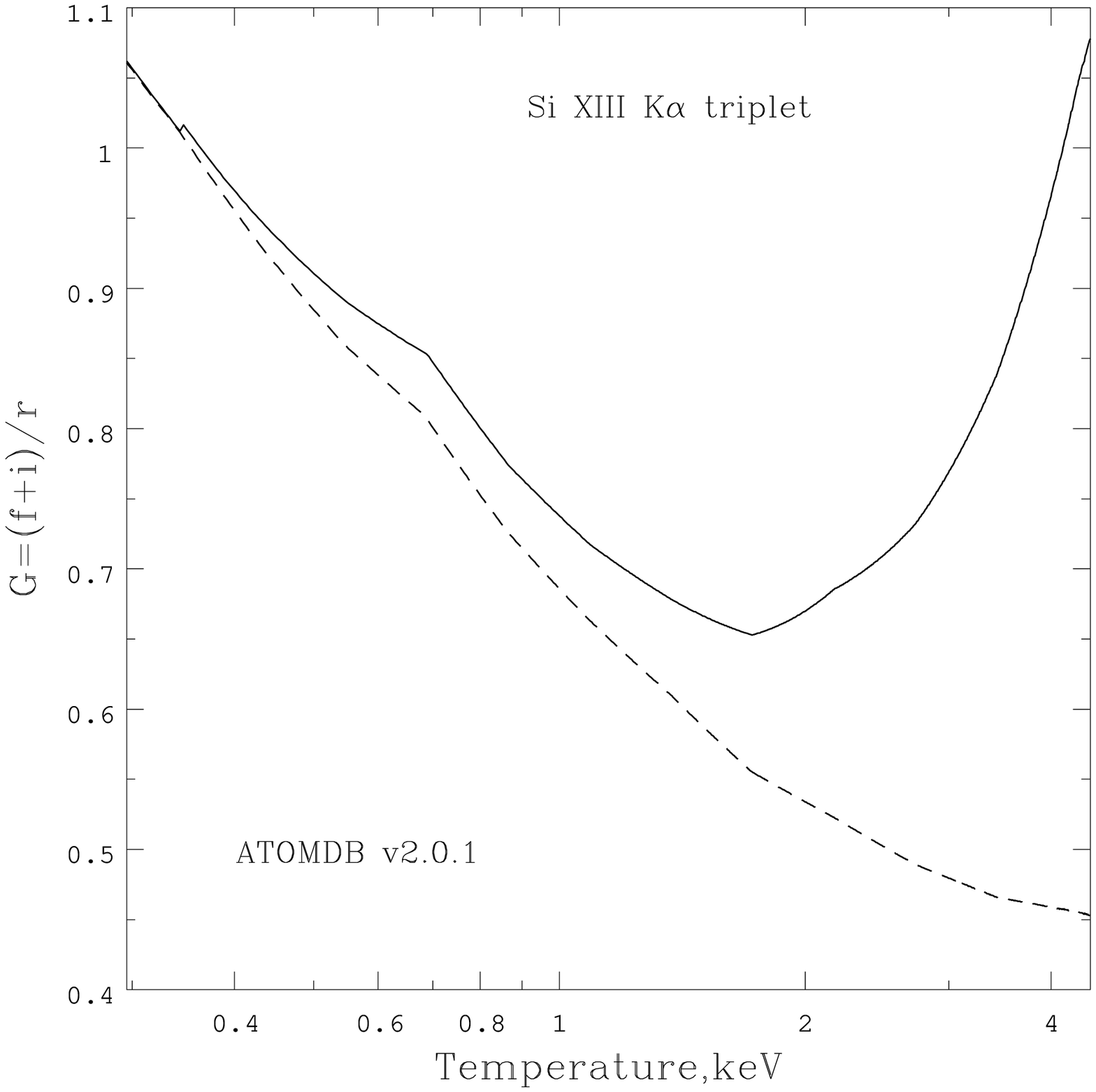}{}{a}}\\
{\centering \leavevmode
\smfigure{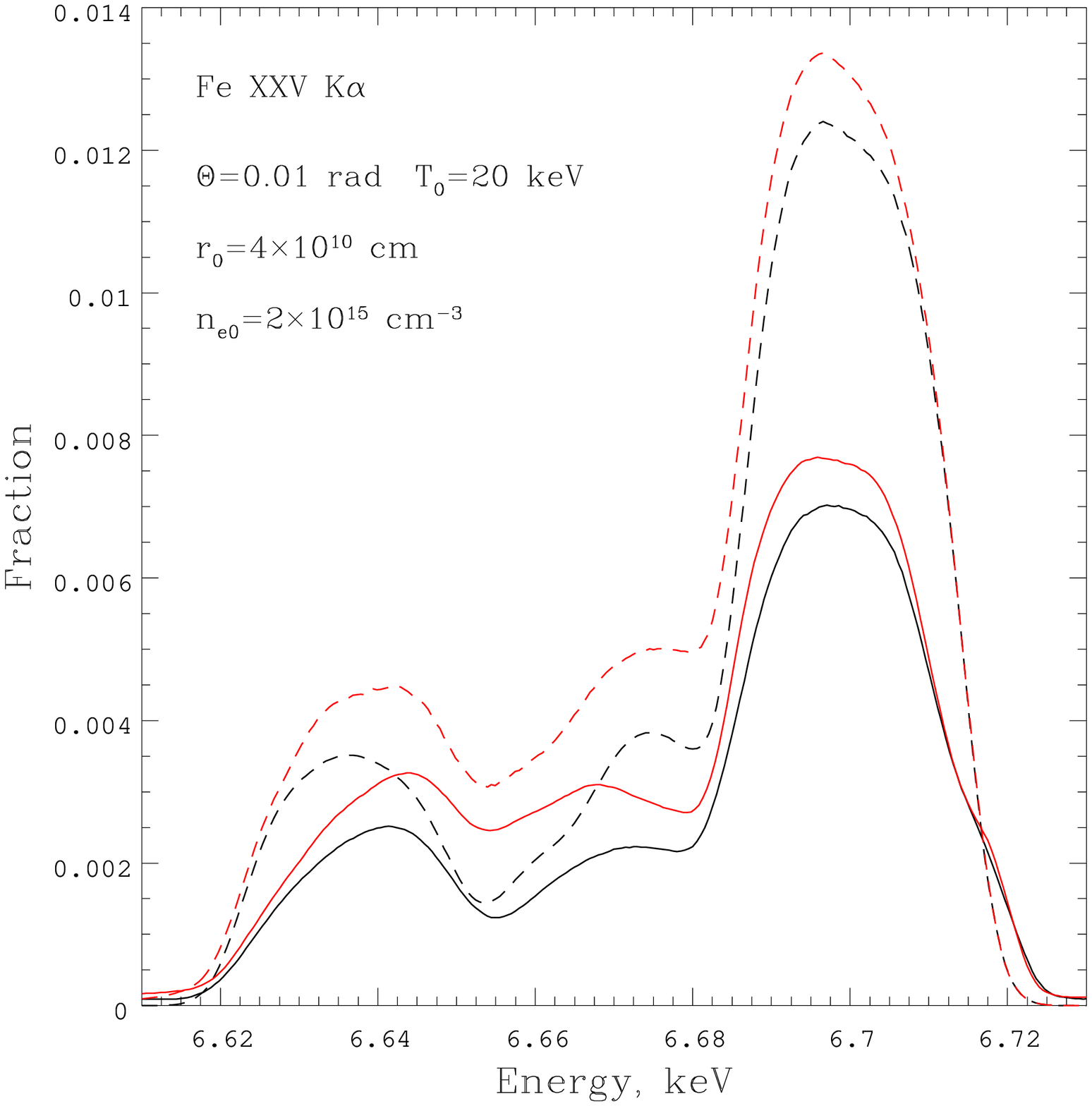}{}{b}}
\caption{(a) Influence of the Mg XII $ Ly_{\gamma} $ line on the ratio G = (f + i)/r in the Si XIII $K_{\alpha}$ triplet: the solid and dashed curves
are for the cases with and without Mg XII $ Ly_{\gamma} $, respectively. The computation was performed using ATOMDB v2.0.1. (b) The
contribution from the satellites of the Fe XXV $K_{\alpha}$ triplet to the radiation intensity at 6.7 keV: the dashed and solid curves are
for the cases without and with scattering, respectively; red--with satellites, black--without satellites.}
\label{satfig}
\end{figure}

For the triplets of helium-like ions, the bulk of
the satellite intensity integrated over the entire jet is
concentrated in about 25 lines inside the triplet, i.e.,
between the forbidden and resonance lines. In this
case, the satellite intensity relative to the resonance
line slightly increases with increasing charge of the
ion nucleus Z. For example, the contribution of
the satellites in the jet model under consideration for
sulfur (Z = 16) is about 12\% of the intensity of the
$K_{\alpha}$ triplet of the helium-like ion and about 20\% for the
same triplet of iron (Z = 26). At the same time, there
can be cases where the energies of the bright lines
corresponding to the transitions in various ions of the
same or different elements are close. In particular,
such a situation takes place for the forbidden line of
the helium-like silicon $K_{\alpha}$ triplet coincident with the
MgXII $Ly_{\gamma}$ doublet. As a result, not only the normalization
of the ratios R and G (see Section \ref{sxvrg}) but
also the shape of the temperature dependence G(T)
itself changes (see Fig. \ref{satfig}a).\

In the case of intrinsically broad lines, the satellites
cannot be resolved in principle, but their contribution
to the total intensity of the blend and its width can be
substantial. For example, for the triplet of helium-like
iron, the effective increase in the width $ W $ of the blend
components (for a definition of $  W$, see Section  \ref{fexxv})
due to the contribution of satellites turns out to be
$ \Delta W \gtrsim 1 $ eV, i.e., $ \gtrsim 5 \% $ at $ \Theta=0.01 $ rad (see Fig. \ref{satfig}b).

%\clearpage
\subsubsection{Nutations}
\
The nutation and, possibly,
jitter of the jets \citep{suzaku10} are other factors
contributing to the broadening of the observed lines.
To estimate the nutation broadening, we used the
model of ephemerides  \citep{katz87} with the parameters found by \cite{gies02} that defines $  z$ for each
of the jets at an arbitrary instant of time. The influence
on the line width can be described in terms of the
total change $\Delta z$ in some time interval $\left(t_{0},t_{0}+t_{obs}\right)$
short compared to the nutation period (6.28 days),
where $t_{0}$ is the starting time of observations and $t_{obs}$
is their total duration. For example, for $t_{obs}=20$ ks,
the maximum value of $\Delta z$ is about 0.0024, which
is in good agreement with the measured maximum
nutation variability $dz/dt\approx$ 0.004 in 0.33 day 
\citep{fabrika04}. In this case, as long as the total time of
observations is much shorter than the nutation period
(6.28 days), $\Delta z$ scales linearly in $t_{obs}$ with a good
accuracy. The change in the energy $E_{0}$ of the line
centroid due to nutation can be estimated as 
$\delta E_{0}=\frac{E_{0}}{(1+z)^{2}} \Delta z$. 
The full width at half maximum of the line
with nutation is then

\begin{equation}
FWHM_{n}\approx\sqrt{FWHM^{2}+\frac{2ln2}{3}\delta E_{0}^{2}},
\label{fwhmn}
\end{equation}
where $ FWHM $ is the full width at half maximum
without nutations. The values of $\Delta z$  and $\delta E_{0}$ for
the Fe XXV (6.7 keV) and Si XIII (1.86 keV) lines
of the ``blue'' jet for specific values of $t_{0}$ and $t_{obs}$
corresponding to the Chandra observations are given
in Table \ref{nutation}.\\

  \begin{table}
\centering
\caption{Nutation broadening of the Fe XXV (6.70 keV)
and Si XIII (1.86 keV) lines for phases corresponding to
the Chandra observations, a report on which is contained
in \cite{marshetal02} [1], \cite{nametal03} [2], and
\cite{lopezetal06} [3].}
\begin{tabular}{lccccc}\hline\hline
\multirow{2}*{Report} &\multirow{2}*{$ z_{b} $}&\multirow{2}*{$t_{obs}$, ks}&\multirow{2}*{$ \Delta z, 10^{-4} $}&\multicolumn{2}{c}{$\delta E_{0}$, eV}\\
{} &{}&{}&{}&{Fe}&{Si}\\
\hline 
{
%\cite{marshetal02}
[1]} & -0.078 &32.0 & 5  & 8 & 2.2 \\
{
%\cite{nametal03}
[2]} &   0.046   &21.3  & 9  & 11 & 3.1 \\
{
%\cite{lopezetal06}
[3]} & 0.014   &25.7  & 10  & 13  & 3.6\\
\end{tabular}
\label{nutation}
\end{table}
    
  Given the observed line widths, it follows from
relation (\ref{fwhmn}) that the nutation broadening can serve
as an additional ($\sim 5\% $) correction to the FWHM.

\subsubsection{Photoionization and photoexcitation}

So far our analysis was based on the assumption that
the atoms in the plasma of the jets in SS 433 were
ionized and excited exclusively by collisionswith electrons;
in this case, the influence of the electromagnetic
radiation fields permeating the jets is negligible.
How valid is this assumption?\

As regards the intrinsic X-ray radiation from the
jets, it should not affect significantly the ionization
balance of the hot plasma, because the jets are optically
thin for the ionizing continuum. However, the
gas of the relativistic jets in SS 433 can be irradiated
by the much more intense X-ray radiation produced
inside the funnel of a thick accretion disk near the
black hole \citep{fabrika04}. Indeed, if SS 433 is
similar to the ultraluminous X-ray sources observed
in nearby galaxies \citep{rob07} but only turned to
us by its edge rather than by the accretion disk plane,
then the X-ray luminosity emitted within the funnels
can be $L_{x}\sim 10^{40}$--$10^{41}$ erg/s (when recalculated
to an isotropic source). The photoionization fraction
of the jet gas in the field of such radiation will be
determined by the ionization parameter $\xi=L/nr^2$
and, depending on the density $n$ and the distance
to the jet base $ r$, can reach $10^3$ or even greater
values. In this case, the influence of photoionization
on the ionization balance in the hot plasma of
the jets can be significant (e.g., \cite{kallman82}). 
However, Chandra HETGS measurements
\citep{marshetal02} showed that the permitted
line dominates in the resolved (in energy) $K\alpha$ triplets,
primarily of SiXIII. This means that collisions dominate
over photoionization in the formation region
of these lines. Since the ionization parameter $\xi=L/nr^2$
 does not change greatly along the X-ray jet for
both quasi-adiabatic and quasi-cylindrical models,
collisions dominate over photoionization during the
excitation of levels in the atoms of heavier elements as
well. Nevertheless, we are going to consider in more
detail the influence of X-ray radiation from the inner
accretion disk regions on the ionization and radiative
properties of the plasma in the jets of SS 433 in a
future paper.\

The supercritical accretion disk in SS 433 is also
a powerful source of ultraviolet radiation with a characteristic
temperature of $\sim 50000$  K and a luminosity
of $\sim 10^{40}$ erg/ s \citep{dolan97}. This is also
confirmed by observations of the radio nebula W50
produced by the impact of the jets in SS 433 on the
interstellar medium \citep{fabrika04}. An ultraviolet
photon of appropriate energy can excite the electron
from the upper level ($^3S$) of the forbidden transition
in a helium-like ion to the upper level ($^3P$) of the
intercombination transition \citep{PD01}. In
a strong ultraviolet radiation field, this effect leads
to a redistribution of the radiation fluxes in the forbidden
and intercombination lines of the triplets in
favor of the latter, i.r., to a decrease in ratio $R$, just
as in the case of a high gas density. This is particularly
true for the lighter elements. Calculations
\citep{PD01} (and our Cloudy computations)
for a blackbody radiation field with a temperature of
50 000 K show that the effect for the SiXIII triplet
must be significant even if the dilution of radiation related
to the fact that the disk radiation permeates the
jets from some distance and at some angle is taken
into account. Since there is a considerable uncertainty
in determining the characteristic temperature
and the geometry of the ultraviolet radiation source
\citep{gies02}, we will restrict ourselves here
only to the above remark of a somewhat qualitative
character.

 %\clearpage 
%%%%%%%%%%%%%%%%%%%%%%%%%%%%%%%%%%%%%%%%%%%
%\newpage
\section{Comparison with observations}

In this part, we would like to discuss the application
of our simulation results to analyzing and
interpreting the Chandra spectra of SS 433 at the
phase of the greatest disk opening toward the observer
\citep{marshetal02}. The weighted mean
line width in these observations corresponded to the
scenario of a ballistic flow with an opening angle
$ \Theta=0.^{\circ}61 \pm 0.^{\circ}03\approx 0.01$ rad. The jet density $ n_{e} \sim 10^{14} $cm $ ^{-3} $ in the line formation region ($ T \sim 1$ keV)
was estimated from the SiXIII $ K_{\alpha} $ triplet. 
The physical volume of this region, hence, the distance of this region from the cone apex (assuming a conical flow with an opening angle$ \Theta$ ) was estimated 
from the emission measure of the component with such a temperature
in a four-temperature model describing satisfactorily
the line intensities. Using the estimates of
the emission measure for the other three components
and assuming the temperature profile to be adiabatic,
 the corresponding locations and densities
of these components were found. As a result,  a
physical picture was obtained that provided satisfactory agreement with the
observed spectrum but, at the same time, had significant
shortcomings. Below, we would like to show
how this picture changes if the scattering effects we
considered and others (see the ``Results'' Section) are
taken into account. Since the triplets of helium-like
iron (FeXXV $ K_{\alpha} $) and silicon (SiXIII $ K_{\alpha} $) were the
main diagnostic tools, we will dwell on a comparison
of themeasured line characteristics with the results of
our calculations.\ 

\begin{figure*}[h!]
\centering
\epsfxsize=1.0\textwidth \epsfbox{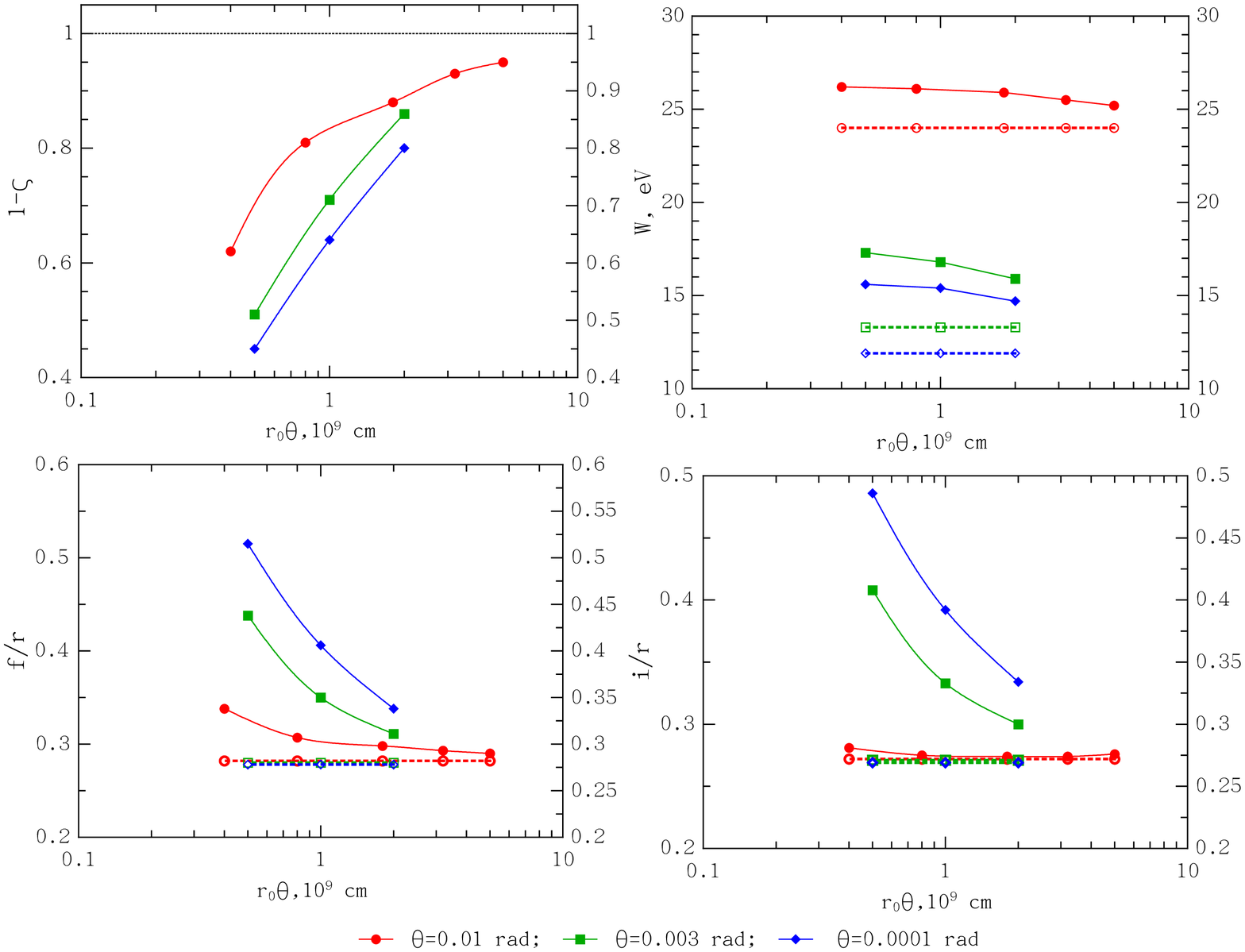}
\caption{Analysis of the simulation results for the FeXXV $ K_{\alpha} $ triplet in the Chandra HEG resolution for the approaching jet
with $ T_{0}=20$ keV at the phase of the greatest disk opening toward the observer. The dashed lines (with the corresponding
open symbols) correspond to the unscattered line characteristics.}
\label{fesum}
\end{figure*}

As has already been noted above, the Chandra
HETGS spectral resolution ($ \sim
200 @ 6.7 $ keV) is insufficient
for investigating the fine structure of the
FeXXV $ K_{\alpha} $ triplet ($ \sim 300 @ 6.7 $ keV is needed). Therefore,
when the observed profile is analyzed, some
fixed ratios of the components (f/r and i/r) is assumed
(e.g., f/r = 0.23 and i/r = 0.28 in \cite{nametal03}). Figures \ref{fesum}a and \ref{fesum}d show how
strongly these ratios are subjected to the scattering
effects, which points to the necessity of correcting
the line widths estimated within a simple model. An
unresolvable triplet with enhanced (compared to the
resonance line) forbidden and intercombination components
can appear as consisting of broader components
with a bright resonance line but with a shifted
centroid. Bearing in mind the broadening by scattering
(Fig. \ref{fesum}b) and the additional possibilities for
broadening from the ``Results'' Section, it should be
recognized that the models with $ \Theta <0.01 $ rad are
admissible from the viewpoint of observations in the
high-energy part of the spectrum. In addition, the
decrease in triplet intensity (Fig. \ref{fesum}a) due to the light
scattering effects implies that the emission measure
of the hottest parts of the jet estimated from the
observed iron line intensities and the heavy-element
abundances estimated from the ratio of the observed
intensities of the corresponding lines should be corrected
significantly. Thus, the scattering effects affect
significantly the jet parameters ($ \Theta, n_{e0}, r_{0} $) estimated
from the FeXXV $ K_{\alpha} $ triplet characteristics in Chandra
observations. \  
  
\begin{figure*}[h!]
\centering
\epsfxsize=1.0\textwidth \epsfbox{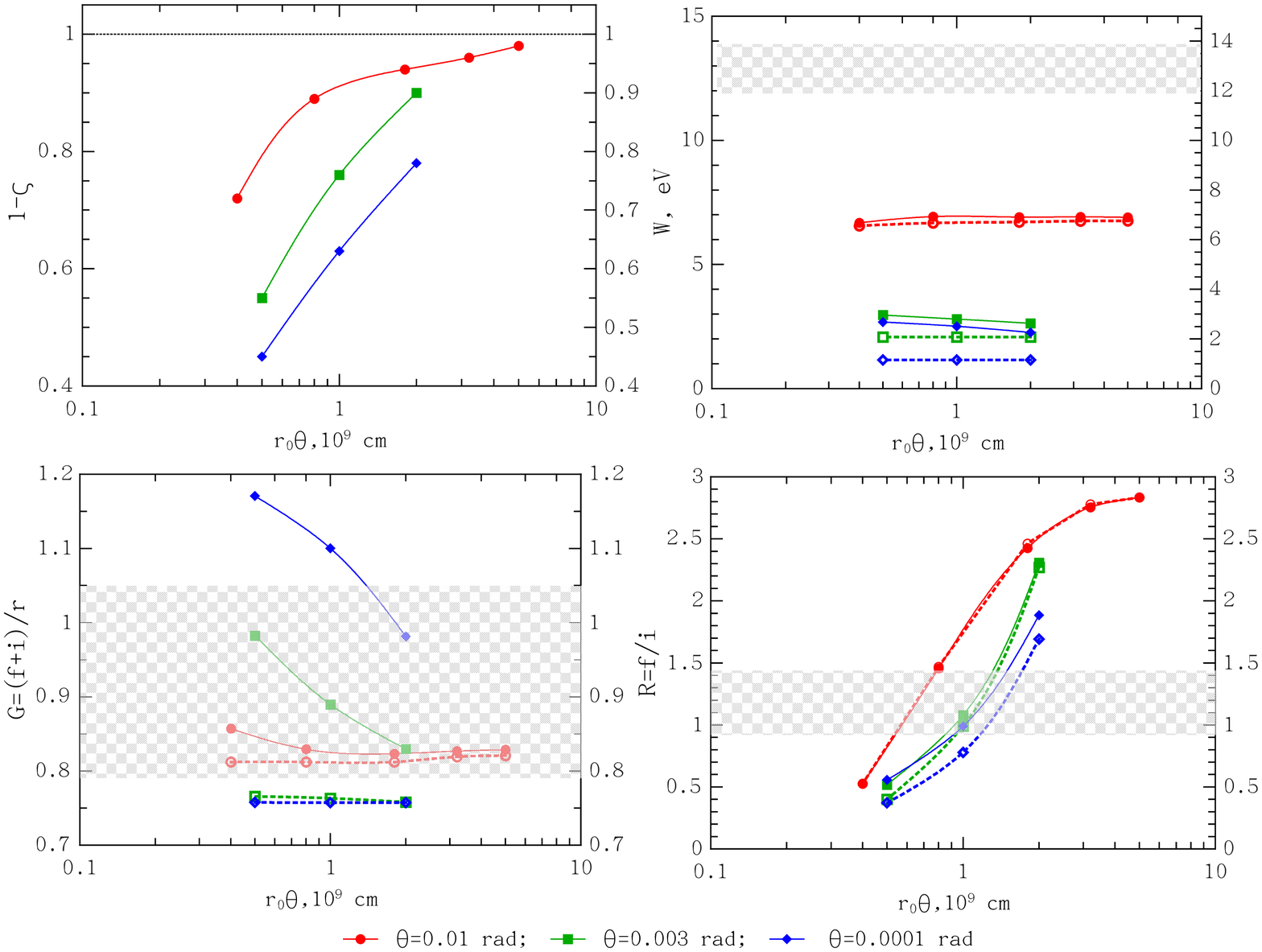}
\caption{Analysis of our simulation results for the SiXIII $ K_{\alpha} $ triplet in the Chandra HEG resolution for the approaching jet
with $ T_{0}=20$  keV at the phase of the greatest disk opening toward the observer. The dashed lines (with the corresponding
open symbols) correspond to the unscattered line characteristics. The ratios R and G are presented by taking into account
the contribution from the MgXII $ Ly_{\gamma}$ doublet. The shaded regions correspond to R and G measured in Chandra observations
\citep{marshetal02}.}
\label{sisum}
\end{figure*}

 In contrast, the remarkable spectral characteristics
of the Chandra instruments in the low-energy region
\textbf{allow} the fine structure of the helium-like silicon
triplet (and, to some extent, the SXV $ K_{\alpha} $ triplet) to be
investigated, but, at the same time, they also require
a more detailed study of their formation mechanisms
for interpreting the observed line characteristics. The
observed width of the SiXIII $
K_{\alpha} $ triplet components
(FWHM $ \approx $1900 km/s) corresponds to a jet opening
angle $ \Theta \simeq 0.015 $ rad
 \footnote{ It is worth noting that in \cite{marshetal02} the weighted
mean width FWHM $ \approx $ 1710 km/s measured from all lines
corresponds to an opening angle $ \Theta\approx 0.^{\circ}77 \approx 0.014 $ rad
(which is easy to see from Eqs. (3) and (4) of this paper),
while the authors provide $ \Theta\approx 0.^{\circ}61\approx 0.01 $ rad.}. 
 This makes the interpretation
of the widths in terms of quasi-cylindricalmodels very
problematic, even if the possible broadening effects
we considered are taken into account (Fig. \ref{sisum}b).
Thus, the models with $ \Theta \geq 0.01 $ rad are much more
preferential. However, the intensity of the SiXIII $ K_{\alpha} $
triplet decreases significantly as a result of scatterings
even in this case (Fig. \ref{sisum}a). This implies the necessity
of a significant correction of the estimated emission
measure for the jet region with $ T \sim 1 $ keV, though
slightly less significant than that for the hotter region
at the jet base. \ 
 
 Figures \ref{sisum}c and \ref{sisum}d present the ratios R and G for
the silicon triplet derived in our calculations by taking
into account the scattering effects and the contribution
from the MgXII $
  Ly_{\gamma} $ doublet, whose energy almost
coincides with the energy of the forbidden silicon
line (see also Section 5.3.1). These results of our calculations
are compared with the ratios (confidence intervals)
R and G measured by \cite{marshetal02}:
$R=  1.18\pm0.26 $ and $G= 0.92 \pm
  0.13 $. The following
conclusions can be drawn from this comparison.
First, the measured value of G is consistent with
most of the quasi-adiabatic and quasi-cylindrical jet
models considered. Second, the measured value of
R points to $ r_{0}\Theta \sim 10^{9} $ cm, which together with
 $ \Theta \simeq 0.01 $ rad estimated from the line broadening gives
$ r_{0} \sim
  10^{11} $ cm. This is consistent with the estimates
from \cite{medved10} but exceeds appreciably
the value obtained by \cite{marshetal02}-
$ r_{0} \sim 2\times 10^{10} $cm, and, at the same time, is less than
$r_{0}\sim 10^{12}$cm found by \cite{kateetal06} from the
eclipse of the jets by the optical companion.

The presented comparison of the results of our
numerical calculations with the results of observations
published by \cite{marshetal02} is more likely
qualitative than quantitative in nature. Obviously,
comparison of the spectral predictions of our model
for the scattering of radiation in the jets of SS 433
directly with the Chandra HETGS data \citep{marshetal02}
 is required to formulate more reliable
conclusions. However, this is beyond the scope of our
paper. \
 
% \clearpage

\section{Conclusions}

Let us briefly list our main conclusions:\

(1) The effect of a decrease in line intensity due
to scattering inside the jet turns out to be very pronounced,
but it does not exceed 60\% in magnitude on
the entire grid of parameters. Thus, such anomalies
as the excess of radiation in the region of the helium-like nickel
triplet are difficult to explain in terms of the scattering
effects alone, it is quite possible to make the problem
less dramatic. At the same time, diagnosing the jet
plasma parameters based on the line intensity ratios
turns out to be impossible without allowance for the
scattering effects, because the introduced shift can
lead to significant systematic errors.\

(2)  The scattering inside the jets, along with the
nutational motion and the contribution of satellites,
lead to a noticeable additional line broadening. This
can lead to overestimates of the opening angle
$  \Theta $ from the line width in Chandra X-ray observations.
As a consequence, the models with $ \Theta \sim
  0.005 $ -- 0.01 rad remain admissible. Therefore, it is
worth noting that the more complex scenario of a
nonconical relativistic flow (for example, ongoing
collimation; \cite{nametal03}), which can be
described by the change in $ \Theta $ along the jet in some
approximation, may actually be realized.\
 
\begin{figure*}[h!]
\centering
\epsfxsize=1.0\textwidth \epsfbox{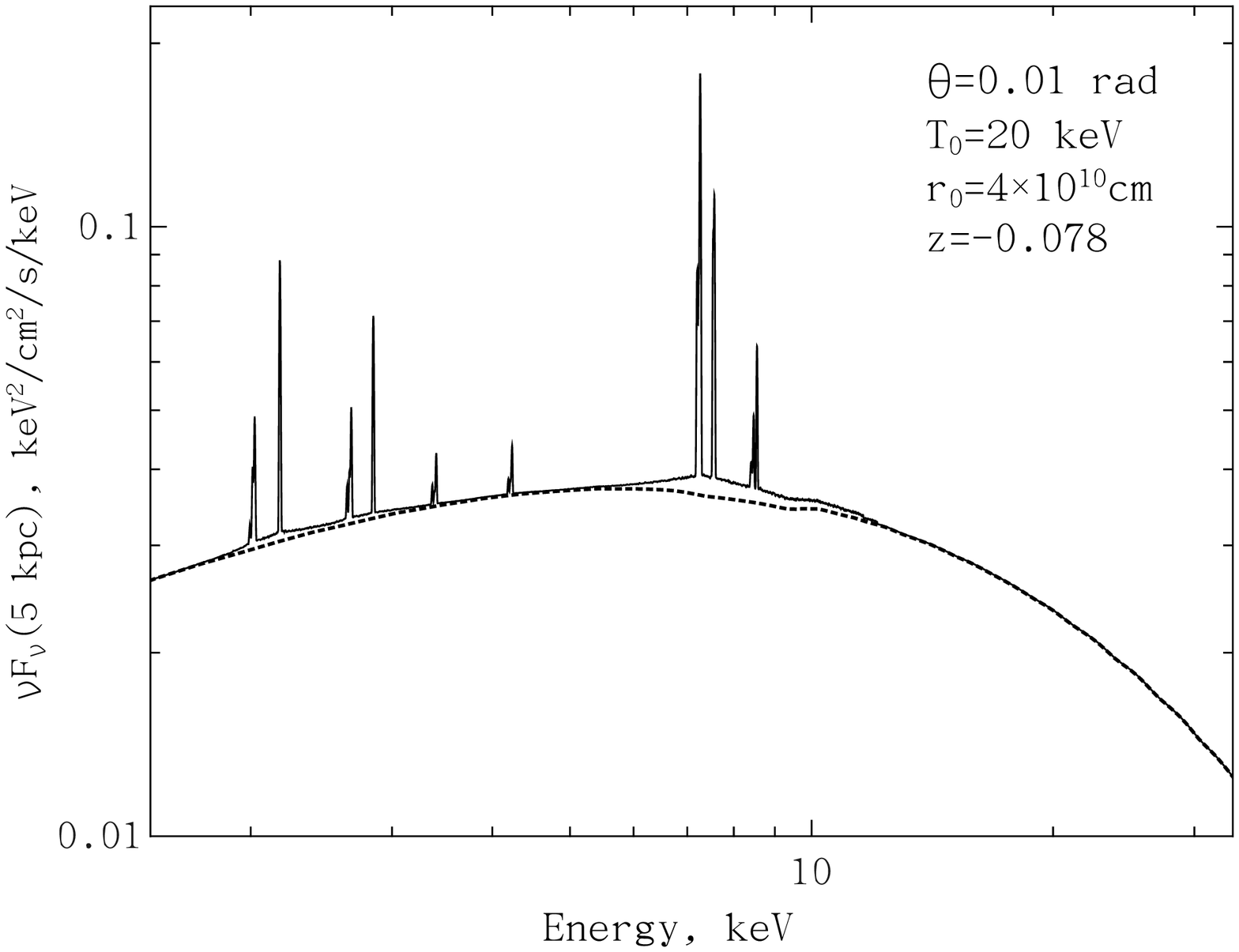}
\epsfxsize=1.0\columnwidth \epsfbox{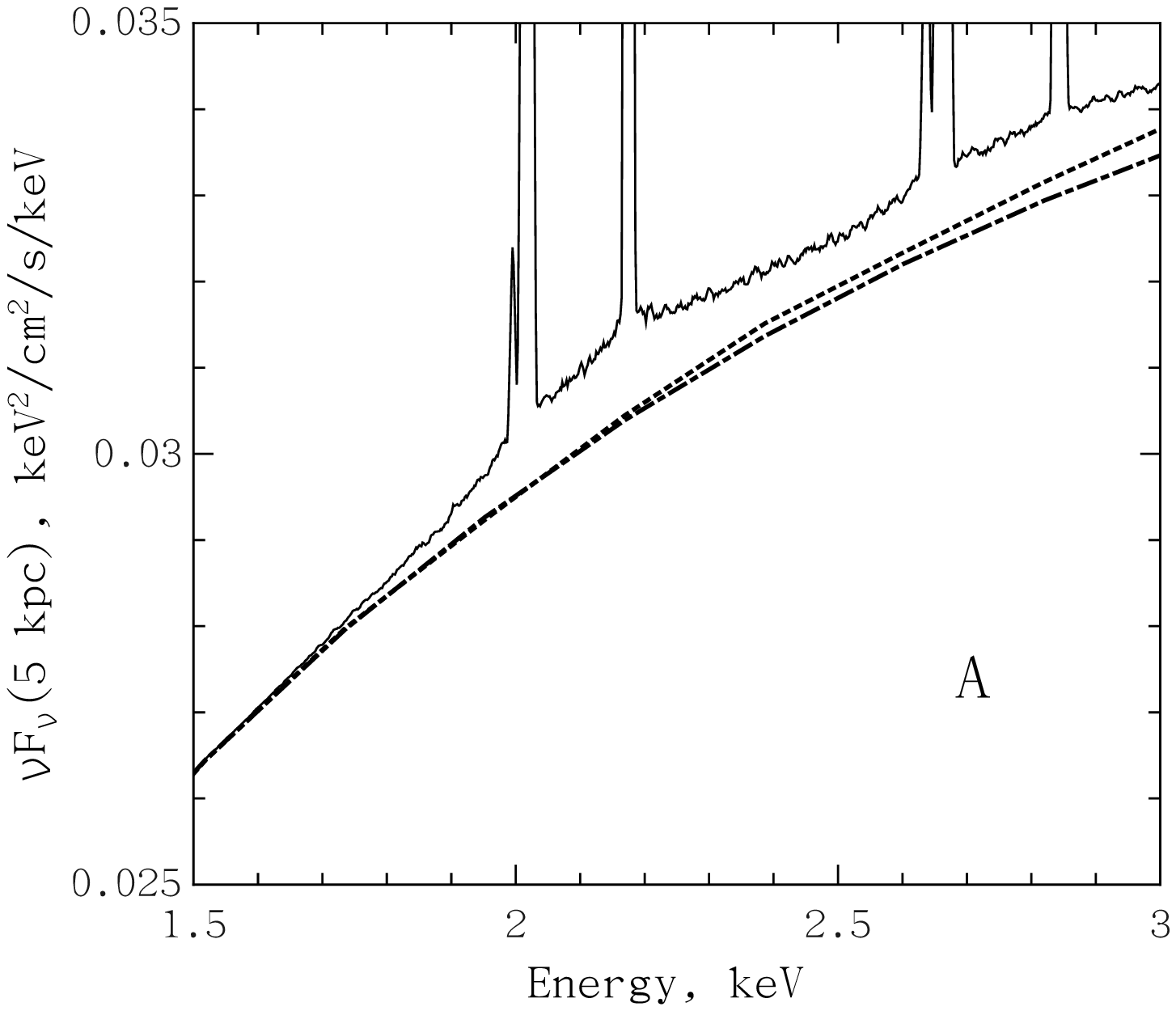}
\epsfxsize=1.0\columnwidth \epsfbox{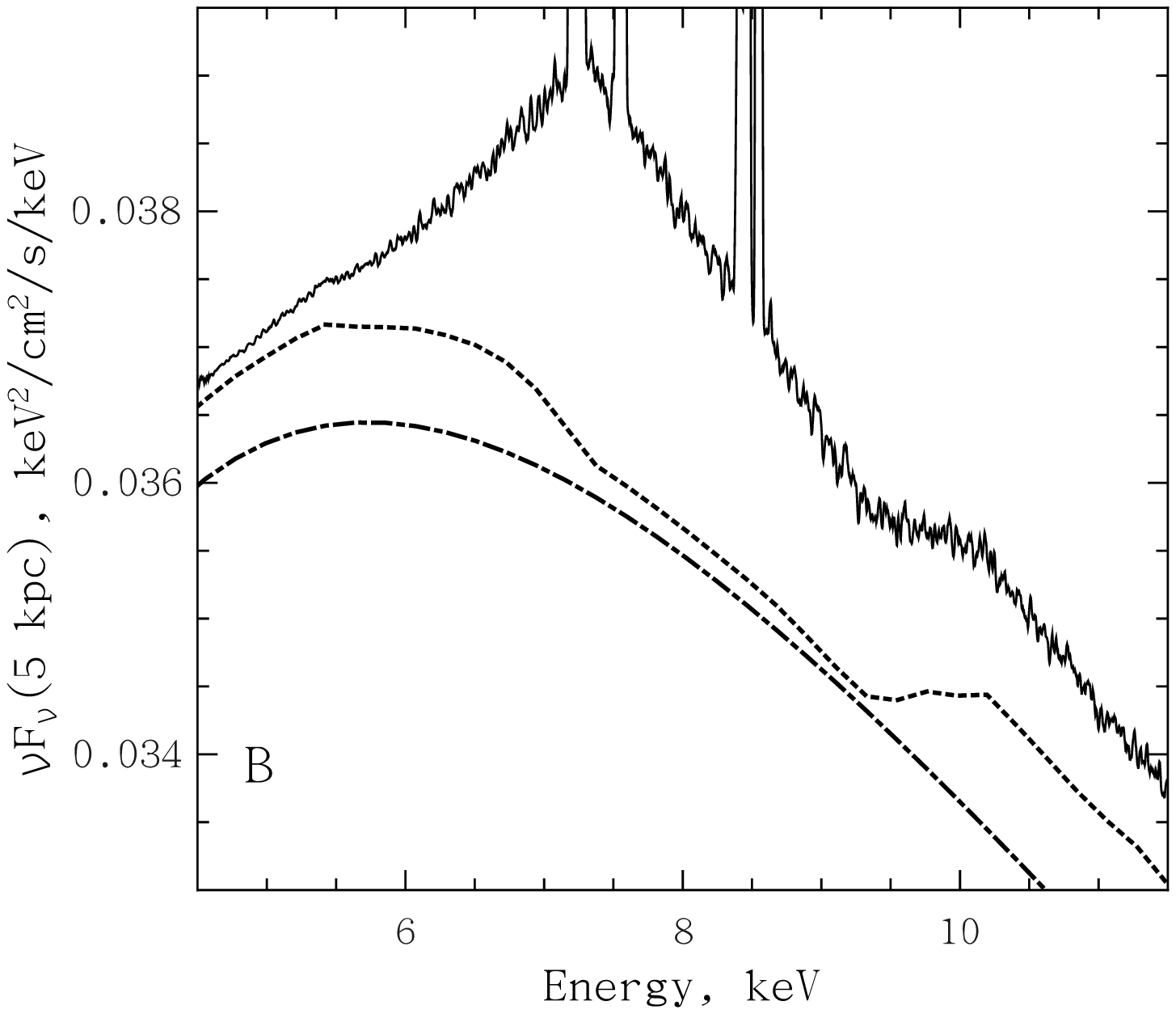}

\caption{ Broadband simulated synthetic spectrum (solid line) of the approaching jet (z = -- 0.078) at a distance of 5 kpc.
The continuum radiation (dashed line) was computed using ATOMDB v.2.0.1 (the NoLine model). All of the lines from Table \ref{linelist}
with allowance made for their scattering inside the jet but without allowance for the contribution of satellites were included.
Magnified fragments of the broadband spectrum in the low-energy (bottom left) and high-energy (bottom right) regions. The dash-dotted line
indicates the contribution from bremsstrahlung (the \textit{brems} model in XSPEC v12.6.0). The difference between the bremsstrahlung and
NoLine continua is due to the presence of recombination and two-photon components.
}
%\vspace*{10mm}
\label{ssth1}
\end{figure*}

(3) There must be broad wings of scattered radiation
near the brightest lines, distorting the shape
of the continuum in a certain way. The magnitude
of this distortions turns out to be of the order of (or
even greater than) the contribution from the recombination
and two-photon components. As an illustration,
we provide a synthetic broadband spectrum
of the approaching jet at a phase corresponding to the
Chandra X-ray observations of SS 433 in 1999 at a
distance of 5 kpc \citep{marshetal02} (Fig. \ref{ssth1}). Accurately
measuring the shape of the scattered component
can give additional information about the density
distribution and the temperature profile along the jet.\ 

(4) The fine structure of the lines is very sensitive
to the scattering effects (especially in the case of
doublets and triplets). This makes its investigation
a powerful tool for diagnosing the jet parameters -- the
density of the emitting region (and, consequently, its
size) and the opening angle $ \Theta $. New-generation X-ray
observatories (primarily Astro-H) equipped with
spectrometers with a resolution of several electronvolts
(microcalorimeters) will allow one to quickly (in
an exposure time of 10 ks, Astro-H will be able to
collect $\sim $600 photons in the FeXXV (6.7 keV) line
and, consequently, $\sim $60 photons in the broad wings)
get a clear idea of the mechanisms for the formation
of X-ray lines in the spectrum of the jets in SS 433.
 
 \bigskip
 \bigskip

\section*{ACKNOWLEDGMENTS}
 \  This work was supported by Programs P-21 and
OFN-16 of the Russian Academy of Sciences and the
Program for Support of Leading Scientific Schools of
the Russian Federation (NSh-5069.2010.2).
%\clearpage
%%%%%%%%%%%%%%%%%%%%%%%%%%%%%%%%%%%%%%%%%%%
%\newpage
%%%%%%%%%%%%%%%%%%%%%%%%%%%%%%%%%%%%%%%%%%%

\end{document}